\documentclass[%
 reprint, 
amsmath,amssymb,
aps,
prx,
superscriptaddress]{revtex4-2}
\usepackage{amsmath}
\usepackage{braket}
\usepackage{graphicx,xcolor}% Include figure files
\usepackage{dcolumn}% Align table columns on decimal point
\usepackage{bm}% bold math
\usepackage[utf8]{inputenc}
\usepackage{url}
\usepackage[thinc]{esdiff}
\usepackage{mathdots}
\usepackage{booktabs}
\usepackage{makecell}
\usepackage{hyperref}

\begin{document}
% \linenumbers

\preprint{APS/123-QED}

\title{Probing the Dynamics of Two-Level System Defect Ensembles via Broadband Cryogenic Transient Dielectric Spectroscopy}

\author{Qianxu Wang}
\affiliation{\mbox{Department of Physics and Astronomy, Dartmouth College, 6127 Wilder Laboratory, Hanover, New Hampshire 03755, USA}}
\affiliation{\mbox{Thayer School of Engineering, Dartmouth College, 15 Thayer Drive, Hanover, New Hampshire 03755, USA}}

\author{Juan S. Salcedo-Gallo}
\affiliation{\mbox{Thayer School of Engineering, Dartmouth College, 15 Thayer Drive, Hanover, New Hampshire 03755, USA}}

\author{Sara Magdalena G{\'o}mez}
\affiliation{\mbox{Department of Physics and Astronomy, Dartmouth College, 6127 Wilder Laboratory, Hanover, New Hampshire 03755, USA}}
\affiliation{\mbox{Thayer School of Engineering, Dartmouth College, 15 Thayer Drive, Hanover, New Hampshire 03755, USA}}

\author{Roy Leibovitz}
\affiliation{\mbox{Department of Physics and Astronomy, Dartmouth College, 6127 Wilder Laboratory, Hanover, New Hampshire 03755, USA}}

\author{Jake Freeman}
\affiliation{\mbox{Department of Physics, Middlebury College, Middlebury, Vermont 05753, USA}}
\affiliation{\mbox{Thayer School of Engineering, Dartmouth College, 15 Thayer Drive, Hanover, New Hampshire 03755, USA}}

\author{Sof{\'i}a {\'A}brego}
\affiliation{\mbox{Department of Physics and Astronomy, Dartmouth College, 6127 Wilder Laboratory, Hanover, New Hampshire 03755, USA}}
\affiliation{\mbox{Thayer School of Engineering, Dartmouth College, 15 Thayer Drive, Hanover, New Hampshire 03755, USA}}

\author{Simon A. Agnew}
\affiliation{\mbox{Thayer School of Engineering, Dartmouth College, 15 Thayer Drive, Hanover, New Hampshire 03755, USA}}

\author{William J. Scheideler}
\affiliation{\mbox{Thayer School of Engineering, Dartmouth College, 15 Thayer Drive, Hanover, New Hampshire 03755, USA}}

\author{Salil Bedkihal}
\affiliation{\mbox{Thayer School of Engineering, Dartmouth College, 15 Thayer Drive, Hanover, New Hampshire 03755, USA}}

\author{Mattias Fitzpatrick}
\email{mattias.w.fitzpatrick@dartmouth.edu}
\affiliation{\mbox{Department of Physics and Astronomy, Dartmouth College, 6127 Wilder Laboratory, Hanover, New Hampshire 03755, USA}}
\affiliation{\mbox{Thayer School of Engineering, Dartmouth College, 15 Thayer Drive, Hanover, New Hampshire 03755, USA}}

\begin{abstract}
Two-level system (TLS) defects in dielectrics are a major source of decoherence in superconducting circuits, yet their microscopic origin and distribution remain poorly understood. Existing circuit-QED probes access limited frequency ranges and mode volumes, restricting studies of isolated materials and interfaces. Here, we present Broadband Cryogenic Transient Dielectric Spectroscopy (BCTDS), a technique for probing TLS-hosting materials over a broad frequency range at cryogenic temperatures. Under strong finite-duration microwave excitation, the transient homodyne I–Q response exhibits coherent phase dynamics after the drive is turned off. Fourier analysis of the transient phase reveals characteristic V-shaped structures that move between cooldowns, consistent with thermocycling-induced changes in the local TLS defect environment that shift defect resonance frequencies. The transient response of BCTDS further enables estimation of susceptibility and two-time correlation functions of the TLS defect ensemble. The observed phase dynamics are qualitatively captured by a driven standard tunneling model containing only a few representative TLS defects. Despite its simplicity relative to the full experimental ensemble, the model reproduces the essential Floquet-dressed dynamics during the drive and generates post-pulse V-shaped structures and interference fringes consistent with the experimental data. The observed BCTDS response may reflect a crossover from localized TLS defect dynamics to a delocalized regime under strong driving, before being quenched into a transient regime that reflects the TLS defect resonance frequencies. Overall, BCTDS represents a potentially useful broadband, time-resolved wafer-level approach for probing TLS defects relevant to quantum technologies.
\end{abstract}

\date{\today}
\maketitle

\section{\label{sec:intro}Introduction}
Many amorphous and disordered materials display low-temperature thermodynamic behavior that is remarkably universal across a wide range of systems. These properties deviate sharply from the predictions of crystalline Debye theory. In particular, at cryogenic temperatures, the specific heat scales approximately linearly with temperature, \( C(T) \propto T \), rather than following the \( T^{3} \) dependence expected for crystalline solids~\cite{Zeller1971}. The recurring appearance of this anomalous scaling in chemically and structurally distinct materials suggests the presence of a common class of low-energy excitations beyond phonons, likely originating from structural or dynamical defects intrinsic to amorphous materials and interfaces.

\begin{figure}[htpb!]
\begin{centering}
\includegraphics[width=0.46\textwidth]{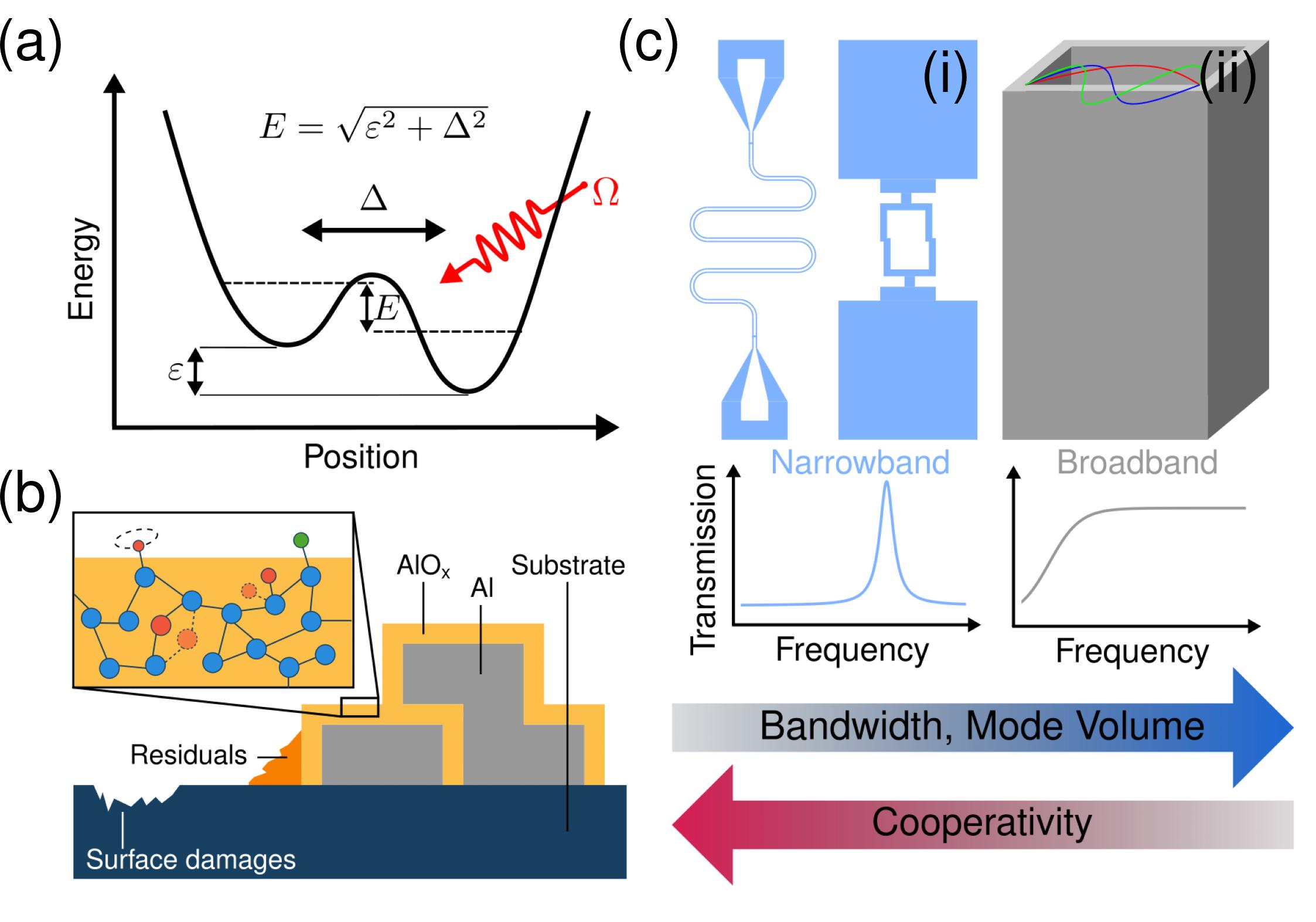}
\end{centering}
\caption{\label{intro} Overview of TLS defects and how they can be probed.
(a) Double-well representation of TLS defects under a periodic drive $\Omega$, described by the standard tunneling model. 
(b) Candidate TLS defects and their likely locations in superconducting circuits.
(c) Comparison of traditional TLS defect spectroscopy using 2D qubits and resonators and our proposed broadband 3D waveguide approach. The concept of image (b) is inspired by \cite{Lisenfeld2019}.}
\label{fig:overview}
\end{figure}

A phenomenological description of these excitations is provided by the standard tunneling model, illustrated in Fig.~\ref{fig:overview}(a), which was introduced independently by Anderson, Halperin, and Varma~\cite{Anderson1972}, and by Phillips~\cite{Phillips1972}. In this framework, a subset of defects is modeled as effective two-level systems (TLS), corresponding to atoms or small groups of atoms tunneling between nearly degenerate configurations. The model successfully accounts for many experimentally observed features of glassy behavior at low temperatures.
Despite this empirical success, the tunneling TLS defect framework falls short of a complete microscopic theory of amorphous solids. In particular, it does not explain the apparent universality of low-temperature properties across chemically and structurally diverse materials, nor does it fully capture the interactions and collective effects expected in realistic disordered systems.

%%%%%%%%%%%%%%%%%%%%%%%%%%%%%%%%%%%%%%%%%%%%%%%%%%%%%%%%%%%%%%%%%%
Although the notion of a TLS defect was initially developed to explain the thermodynamic properties of glassy materials, such defects have recently regained relevance in solid-state quantum computing technologies. This renewed interest is especially evident in superconducting circuits, which have become a leading platform for quantum computing primarily due to their fast gate times, design flexibility, and compatibility with advanced semiconductor fabrication techniques. However, their decoherence times are relatively short compared to other promising quantum technologies, such as neutral atoms, trapped ions, and spin qubits. From measurements of bulk loss tangents~\cite{Wang2015, read_precision_2023} and modeling of the Purcell effect~\cite{pozar, reed_fast_2010, purcell_resonance_1946}, it has become clear that there are other sources of loss. Furthermore, measurements of qubit coherence under varying electric fields and mechanical strain provide evidence for TLS defects in superconducting circuits.~\cite{Lisenfeld2010, Grabovskij2012, Lisenfeld2015, Lisenfeld2019, yu_experimentally_2022}. Currently, the prevailing understanding is that decoherence in superconducting circuits is dominated by TLS defects on surfaces and interfaces, as illustrated in Fig.~\ref{fig:overview}(b), although their specific atomistic origin remains unclear~\cite{MacCabe2020, Crowley2023, Chiappina2023, Chen2024, bland2025}.

TLS defects are typically probed via photonic or acoustic resonators~\cite{Gao2008, Calusine2018, Woods2019, McRae2020, PhysRevApplied.17.034025} and qubits~\cite{Wang2015, Martinis2005, Lisenfeld2010, Grabovskij2012, Lisenfeld2015, Lisenfeld2019}. These approaches provide high cooperativity but preferentially detect TLS defects with dipole moments aligned with the directly integrated probes and are restricted to small bandwidths and mode volumes, as depicted in Fig.~\ref{fig:overview}(c). Despite these limitations, several studies have demonstrated the existence of TLS-TLS interactions~\cite{Burnett2014, Mueller2015, Lisenfeld2015}, and non-Markovian dynamics~\cite{Agarwal2024, Odeh2025}, leading to questions about the nature of TLS defect interactions in the absence of strongly coupled probes~\cite{Spiecker2024}.

In addition to their impact on the decoherence of superconducting qubits, TLS defects also influence the fidelity of qubit readout. In particular, TLS defects can manifest in Josephson traveling wave parametric amplifiers (JTWPAs), which are crucial for high-fidelity qubit readout due to their quantum-limited amplification and several GHz of bandwidth~\cite{Macklin2015, Obrien2014}. However, recent experiments show that high-power pulses can excite long-lived dielectric echoes in JTWPAs, which are attributed to ensembles of TLS defects~\cite{boselli2025, delattre_quantitative_2025}. These echoes impair the readout fidelity of experiments utilizing JTWPAs (and likely other parametric amplifiers), motivating careful characterization of the transient dynamics of interacting ensembles of TLS defects. 

Conventional approaches to probing TLS defects using resonators or qubits also require fully fabricated devices, which are resource-intensive and hinder rapid exploration of materials and fabrication protocols. Since TLS defects reside in intrinsically disordered host materials and are active at dilution temperatures, it is natural to ask whether they can be interrogated directly in cryogenic operation. In conventional qubit experiments, microwave fields are typically kept near the single-photon level, so TLS defects are probed mainly as part of a passive glassy environment. As a result, their driven non-equilibrium dynamics remain largely inaccessible. By contrast, the application of a strong monochromatic drive for a finite duration can place TLS defects in a highly non-equilibrium regime. During the drive, the TLS defect states become strongly dressed by the periodic field, giving rise to Floquet-like behavior in which the system evolves according to an effective quasienergy structure generated during the drive. In this regime, transitions that would be forbidden at weak driving become allowed through multi-photon processes, and the TLS defect could display a strongly nonlinear response. Once the drive is turned off, TLS defect relaxes toward equilibrium, revealing its intrinsic relaxation time. Large drives therefore provide a direct window into the actively driven dynamics of TLS defects, exposing phenomena inaccessible in the weak-driving, linear regime and framing the problem as the electrodynamics of a disordered, interacting many-body system subjected to a time-dependent external field.

In this work, we present broadband cryogenic transient dielectric spectroscopy (BCTDS), develop its operating principle, and use it to probe interference effects in strongly driven ensembles of TLS defects in isolated materials. BCTDS is a modular, non-invasive technique based on a rectangular broadband waveguide, where finite-duration microwave pulses drive the sample, and the emitted post-pulse field is measured by homodyne detection. A related application of this platform, including coherent-control measurements and the sensitivity of the transient response to thermal cycling, has been reported in our earlier work Ref.~\cite{MQT}. Here, we focus on the physical mechanism of BCTDS and on the transient response of strongly driven disordered TLS defect ensembles. BCTDS enables systematic studies of material-specific TLS defect properties and dynamics, providing insights that can inform the design of next-generation quantum hardware~\cite{megrant_scaling_2025}. We combine experiments with driven standard tunneling model simulations and analytical modeling to show how finite microwave pulses generate phase-resolved spectral features, pulse-duration-dependent interference, and temporal correlations in the emitted field. TLS defect ensembles under a drive can exhibit transient collective dressed-state dynamics analogous to the Mollow triplet of a single TLS defect~\cite{PhysRev.188.1969}. Unlike conventional Electron Spin Resonance (ESR), BCTDS probes the broadband transient electromagnetic and dielectric response following microwave excitation.

Throughout the manuscript, we distinguish between directly measured quantities, model-based interpretations, and future quantitative goals. The directly measured quantities are the transient homodyne quadratures, their amplitude and phase, pulse-duration dependence, and time-domain correlations. The model-based interpretation connects the phase-FFT features and interference fringes to detuning-dependent phase evolution in a driven TLS defect ensemble. Quantitative extraction of TLS defect density, microscopic defect identity, or low-power resonator loss remains a future goal that will require calibrated field participation and differential background subtraction. 

At low temperatures, TLS defect ensembles in disordered materials are strongly localized due to the broad distribution of their energy splittings. In the absence of driving, most TLS defects are far detuned from one another, and their interactions are dominated by longitudinal (ZZ-type) couplings, which primarily produce static energy shifts rather than the exchange of excitations. As a result, coherent transport is suppressed, and the ensemble behaves as a collection of weakly hybridized localized defects.
Applying a strong monochromatic microwave drive for a finite duration brings the TLS defect ensemble into a dressed regime, where each defect is described by Floquet states incorporating drive-induced Stark shifts. TLS defects that are far off-resonant in the bare basis are nevertheless activated by the periodic modulation, which sweeps their dressed energies through avoided crossings during each drive cycle and enables repeated Landau--Zener transitions between dressed states, giving rise to quantum interference effects \cite{SHEVCHENKO20101}. Each passage produces partial hybridization between dressed states, while between crossings the components accumulate a relative dynamical phase set by the instantaneous quasienergy splitting. Over many cycles, these Landau--Zener events generate Stückelberg interference governed by the accumulated phase, with constructive interference occurring when a resonance condition equivalent to the Floquet multiphoton relation $\omega_{0} \approx n \omega_{d}$ is satisfied. Although each off-resonant TLS defect contributes only a small amplitude per passage, coherent accumulation over the pulse duration produces a measurable collective response from many TLS defects.

After the cessation of the drive, the dressed states are no longer eigenstates, and the accumulated superpositions are projected onto the bare TLS basis. The relative phase built up during the driven evolution is thereby mapped onto the phase of the emitted field during the ring-down. Because the measurement is performed using homodyne detection, the rapid free evolution at the TLS defect frequency is removed, and the measured phase directly reflects the accumulated phase offset generated during the pulse. As a result, this phase grows linearly, leading to V-shaped features in its Fourier transform, where the base identifies the resonance frequency of a given defect. Counting these phase V-shaped structures may provide a measure of the spectral density of TLS defects within the probed frequency band.

The broadband waveguide is essential in this process, as it captures the full multi-frequency emission generated by the driven TLS defect ensemble, including harmonics and sidebands. Tracing the arms of a given V as a function of pulse duration further reveals interference fringes, demonstrating coherent phase evolution built up during the driven interval. In this way, phase-resolved FFT analysis of the I–Q signal provides a direct window into how a driven TLS defect ensemble stores and releases phase coherence, captures the contribution of off-resonantly driven TLS defects, and reveals signatures of stronger Floquet hybridization mediated by repeated Landau–Zener transitions between dressed states.

\section{Operating Principle}
\label{sec:op_principle}

\begin{figure*}[ht!]
\begin{centering}
\includegraphics[width=0.98\textwidth]{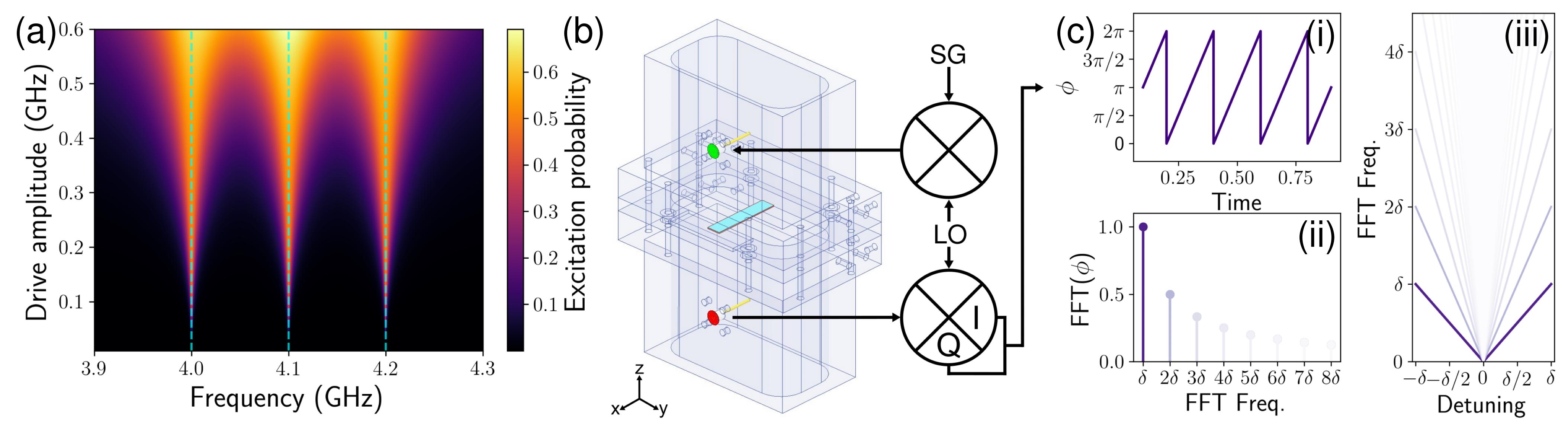}
\end{centering}
\caption{
(a) Excitation probability of three driven TLS defects, showing broadening of the excitation as the drive amplitude increases. (b) Schematic of the BCTDS waveguide and homodyne measurement setup. The waveguide is mounted below the mixing-chamber plate of a dilution refrigerator, while the control and readout electronics remain at room temperature. (c-i) Phase evolution at a rate set by the detuning, producing a wrapped, sawtooth-like phase trace. (c-ii) FFT of the wrapped phase in (c-i), showing Fourier components at integer multiples of the detuning. (c-iii) FFT of phase under a frequency (detuning) sweep, showing nested V-shaped patterns.
}
\label{fig:op_principle}
\end{figure*}

% Chris's para
%To investigate ensemble TLS defect behavior in disordered systems, we begin with a minimal model for a collection of TLS defects, described by the Hamiltonian
%\begin{equation}
%H(t) = \sum_{j}\left[\frac{\Delta_j}{2}\sigma_z^{(j)} + \Omega \cos(\omega_d t)\sigma_x^{(j)}\right],
%\end{equation}
%where $\Delta_j$ is the transition frequency of the $j$th TLS defect, and $\Omega$ and $\omega_d$ are the drive amplitude and frequency, respectively. The first term describes the energy splitting in the TLS basis, while the second term describes the transverse electric-dipole coupling to an external microwave drive. TLS defects in amorphous materials are typically localized and weakly coupled to the electromagnetic field. As a result, resolving their coherent response often requires stronger driving than is used in conventional weak-probe measurements. 
% Modified Para
To investigate ensemble TLS defect behavior in disordered systems, we begin with a minimal model for a collection of noninteracting TLS defects described by
\begin{equation}
H(t)=
\sum_{j}
\left[
\frac{\omega_j}{2}\sigma_z^{(j)}
+
\Omega\cos(\omega_d t)\sigma_x^{(j)}
\right],
\end{equation}
where $\omega_j$ is the transition frequency of the $j$th TLS defect, while $\Omega$ and $\omega_d$ denote the amplitude and frequency of the external microwave drive, respectively. The first term describes the intrinsic energy splitting of each TLS defect, and the second term represents the transverse electric-dipole coupling to the applied drive field. Because of structural disorder in amorphous materials, the TLS defect ensemble is strongly inhomogeneously broadened, such that different defects possess different transition frequencies. In the weak-drive regime, only TLS defects lying within a narrow resonance window around the drive frequency are appreciably excited, leading to effectively localized and approximately independent defect dynamics. As the drive strength increases, the resonance window broadens and progressively larger subsets of TLS defects participate in the driven dynamics, giving rise to nonlinear and drive-dressed ensemble behavior.

To examine TLS defect dynamics in this relatively strong-drive regime, we solve the time-dependent Schrödinger equation in the $\{\ket{e},\ket{g}\}$ basis. This treatment reduces the driven TLS problem to a second-order differential equation with periodic coefficients, equivalent to a generalized Mathieu equation, whose full derivation can be found in Appendix~\ref{app:analytics}. In this approach, the periodic drive naturally generates harmonic components in the TLS defect response. The corresponding Fourier coefficients are determined by a recurrence relation and can be expressed in terms of Bessel functions. The resulting solution gives the total excitation probability for an ensemble of TLS defects \cite{Yingying2024}:
\begin{equation}
P_{\mathrm{total}}(\Omega, \omega_d) =
\sum_{j=1}^{N} \sum_{n}
\frac{1}{2}
\frac{\Omega^2 J_n^2\!\left(\frac{\Omega}{\omega_d}\right)}
{\delta_{j,n}^2 + \Omega^2 J_n^2\!\left(\frac{\Omega}{\omega_d}\right)} ,
\end{equation}
where
\begin{equation}
\delta_{j,n}
=
\omega_{j} - n\omega_d
\end{equation}
is the effective detuning between the $j$th TLS defect and the $n$th drive-induced harmonic.
This expression shows that the periodic drive creates a series of effective resonances, weighted by the corresponding Bessel factor $J_n(\Omega/\omega_d)$. A natural way to interpret this harmonic structure is through the Floquet framework, as described in Appendix~\ref{app:floquet_formalism}. In this picture, the periodic drive dresses the TLS defect and produces a quasienergy spectrum with repeated avoided crossings. At higher drive powers, the dressed states are swept through these avoided crossings, producing repeated Landau--Zener transitions that allow off-resonant TLS defects to participate through multiphoton processes. The same Floquet picture also provides a qualitative mechanism for long post-pulse transients: when the quasienergy spectrum develops small gaps, the corresponding dressed-state superpositions evolve slowly after the pulse, producing longer-lived ring-down features. This connection is illustrated in Fig.~\ref{fig:floquet_quasienergies} of Appendix~\ref{app:floquet_formalism}. Figure~\ref{fig:op_principle}(a) gives the complementary excitation picture for an example ensemble of three TLS defects: as the drive amplitude increases, the resonance window broadens, allowing more TLS defects to participate in the collective driven response.

In conventional resonant measurements, much of this ensemble dynamics remains inaccessible. These measurements are performed with low powers and are most sensitive to individual defects near a selected mode frequency. As a result, they sample only a few strongly coupled TLS defects, while weakly coupled background defects often remain unresolved. In contrast, a broadband geometry with stronger driving can support coherent excitation and detection over a wide microwave band, allowing many TLS defects with different detunings to contribute to the measured response.

Motivated by this picture, we designed a broadband waveguide experiment based on two WR-229-to-coaxial adapters joined by a sample clamp, as shown in Fig.~\ref{fig:op_principle}(b). The sample is inserted through a slot in the clamp and placed at the center of the waveguide. This modular mount allows us to measure samples with different sizes, shapes, and material compositions over a broad microwave bandwidth. In principle, BCTDS can be performed on any sample, although metals and superconductors will alter the boundary conditions of the waveguide, thus altering the electromagnetic modes. However, such a mode can still have spatial overlap with TLS defects in metallic oxides and other materials, meaning that key signatures of TLS defects and other high-Q electromagnetic modes can be observed, including in reflection mode, as demonstrated in Appendix~\ref{app:T_R}. The same approach can also be adapted to patterned films and superconducting devices, where BCTDS could probe TLS defect-sensitive transient responses in more device-relevant geometries. Here, we focus on the dielectric response of TLS defects, which emerges most clearly at cryogenic temperatures. We therefore mount the waveguide assembly below the mixing chamber plate of a Bluefors LD400 dilution refrigerator and perform the experiments at $\sim$10~mK. Across different experiments, we use relatively low drive-line attenuation to deliver strong microwave pulses to the waveguide. Accounting for amplifier gain, attenuation, and cable loss, we estimate the power at the waveguide input to be approximately $-35~\mathrm{dBm}$. This drive excites the sample through the propagating waveguide mode. Details of the fridge wiring and microwave circuitry are given in Appendix~\ref{app:measurement_setup}.

To probe the driven dielectric response, we generate square microwave pulses in the 3--5~GHz band, similar to those used for superconducting-qubit gate operations. The pulses are synthesized directly on an FPGA board (AMD RFSoC 4x2) using the Quantum Instrumentation Control Kit (QICK) software package~\cite{Stefanazzi2022}. The lower end of the frequency range is set by the waveguide cutoff at approximately 3~GHz, while the upper end is set by the validated operating range of the RFSoC readout chain.

When the dielectric sample is driven by a coherent microwave field, the field induces a time-dependent polarization that encodes the absorptive and dispersive response of the sample:
\begin{equation}
    \mathbf{P}(\mathbf{r},\omega)
    \approx
    \varepsilon_0
    \bm{\chi}(\mathbf{r},\omega)
    \mathbf{E}_{\mathrm{loc}}(\mathbf{r},\omega).
\end{equation}
The internal driven quantum dynamics modifies the polarization, and through input-output theory, the resulting emitted microwave field has homodyne quadratures proportional to the two orthogonal components of the TLS defect coherence,
\begin{align*}
I(t) &\propto \langle \sigma_-(t) + \sigma_+(t) \rangle = \langle \sigma_x(t) \rangle, \\
Q(t) &\propto \langle -i(\sigma_-(t) - \sigma_+(t)) \rangle = \langle \sigma_y(t) \rangle .
\end{align*}
from which we can reconstruct the instantaneous amplitude and phase:
\begin{align}
A_{IQ}(t) &= \sqrt{I^2(t) + Q^2(t)}, \\
\phi_{IQ}(t) &= \arg\!\big(I(t) + iQ(t)\big).
\end{align}

As motivated above, under a drive that is stronger than the frequency detuning (Fig.~\ref{fig:op_principle}a), TLS defects become dressed, and their excitation contains a sum over Floquet components. As derived in Appendix~\ref{app:analytics}, the instantaneous TLS defect coherence can be written as the off-diagonal density-matrix element,
\begin{equation}
{\rho_{\mathrm{ge}}}^{(j)}(t)
=
\langle \sigma_+^{(j)}(t)\rangle
\propto
\sum_{n=-\infty}^{\infty}
J_n\!\left(\frac{\Omega}{\omega_d}\right)
e^{-i(\omega_j-n\omega_d)t}.
\end{equation}
for the j-th TLS defect, each component therefore carries a phase
\begin{equation}
\phi_{j,n}(t)
=
\arg\!\left[
e^{-i(\omega_j-n\omega_d)t}
\right]
=
(\omega_j-n\omega_d)t
=
\delta_{j,n}t,
\end{equation}
which is measured by the phase extracted from the IQ signal, up to a constant phase offset relative to the local oscillator and phase delay set by the homodyne detection circuitry:
\begin{equation}
\phi_{IQ}(t)
=
\phi_0+\arg\!\left[
e^{-i(\omega_j-n\omega_d)t}
\right]
=
\phi_0+\delta_{j,n}t.
\end{equation}
This phase evolves linearly in time at the effective detuning. Which, when taken modulo $2\pi$, appears as a wrapped (sawtooth-like) phase evolution (Fig.~\ref{fig:op_principle}(ci)). The sawtooth phase is a periodic function with period $T = {2\pi}/{\delta_{j,n}}$, and admits a Fourier series expansion
\begin{equation}
\phi_{IQ}(t)
=
\frac{a_0}{2}
+
\sum_{n=1}^{\infty}
\left[
a_n \cos(n\delta_{j,n} t)
+
b_n \sin(n\delta_{j,n} t)
\right],
\end{equation}
over one period $0\le t < T$, with coefficients
\begin{align}
a_0 &= \frac{2}{T}\int_0^T \phi_{IQ}(t)\,dt, \\
a_n &= \frac{2}{T}\int_0^T \phi_{IQ}(t)\cos(n\delta_{j,n} t)\,dt, \\
b_n &= \frac{2}{T}\int_0^T \phi_{IQ}(t)\sin(n\delta_{j,n} t)\,dt.
\end{align}
Evaluating these integrals yields
\begin{equation}
\phi_{IQ}(t)
=
\phi_0+\pi
-
2\sum_{n=1}^{\infty}\frac{1}{n}\sin(n\delta_{j,n} t).
\end{equation}
This result shows that a linearly drifting phase produces Fourier components at integer multiples of the detuning,
$|\delta_{j,n}|, 2|\delta_{j,n}|, 3|\delta_{j,n}|,\ldots$,
with decreasing amplitude, as shown in Fig.~\ref{fig:op_principle}(cii). As the drive frequency is swept, the detuning changes accordingly. These Fourier components therefore trace out nested V-shaped structures in the frequency-domain representation, centered at the zero-detuning condition, as shown in Fig.~\ref{fig:op_principle}(ciii). More generally, the phase of the full coherence is given by
$\phi(t)=\arg[\sum_j {\rho_{\mathrm{ge}}}^{(j)}(t)]$, where the coherent sum over all defects and drive-induced harmonics can be written as (see Appendix~\ref{app:analytics} for derivation)
\begin{equation}
\phi(t)
=
\arg\!\left[
\sum_{j}\sum_{n=-\infty}^{\infty}
J_n\!\left(\frac{\Omega}{\omega_d}\right)
e^{-i\delta_{j,n}t}
\right].
\end{equation}
The measured phase response, therefore, contains a superposition of V-shaped features associated with different defect frequencies and harmonic sidebands. These V features provide a frequency-domain signature of the collective TLS defect spectrum encoded in the transient homodyne phase, as demonstrated in the results section below (Fig.~\ref{fig:pulse_width}).

Beyond this phase-based spectral signature, the same transient homodyne measurement also contains information about the complex dielectric response of the TLS defect ensemble. Microscopically, the microwave drive couples to the collective polarization of the TLS defects, and the emitted field measured in homodyne detection is proportional to the resulting time-dependent dipole response. In the weak-drive limit, this connection can be described using the Kubo susceptibility, which relates the induced polarization to the applied field and separates the response into absorptive and dispersive components. Although the strong pulses used in BCTDS can drive the ensemble beyond the strictly linear-response regime, this susceptibility picture provides a useful framework for interpreting the measured IQ transients as an effective dielectric response. Following Novotny \emph{et al.}~\cite{Novotny_2022}, we estimate the dissipative component of the response from the experimentally measured intensity $\mathcal{I}(t)\propto A_{IQ}^2(t)$ as
\begin{equation}
\chi''(\omega)
\propto
\mathrm{Im}
\int_{-\infty}^{\infty}
dt\,
e^{i\omega t}
\left\langle
\mathcal{I}(t)\mathcal{I}(0)
\right\rangle .
\label{eq:chi_imag_IQ}
\end{equation}
A detailed derivation of the connection between the polarization operator, Kubo susceptibility, input-output theory, and the experimentally measured IQ signal is given in Appendix~\ref{app:tls_response}. Together, the phase-V map and the extracted susceptibility provide complementary views of the same driven ensemble response: the phase dynamics identify TLS defect-sensitive response frequencies, while the susceptibility captures the associated frequency-dependent absorption. Because the measured field is emitted into a broadband waveguide, BCTDS probes the coupled response of the driven defect ensemble and its microwave environment. We therefore use temperature, material, pulse duration, and phase-resolved controls to identify features consistent with TLS defect-mediated dynamics. These two analyses form the basis for the experimental results presented below.

\section{Experimental Results}

\begin{figure*}[htpb]
% \begin{centering}
\includegraphics[width=0.98\textwidth]{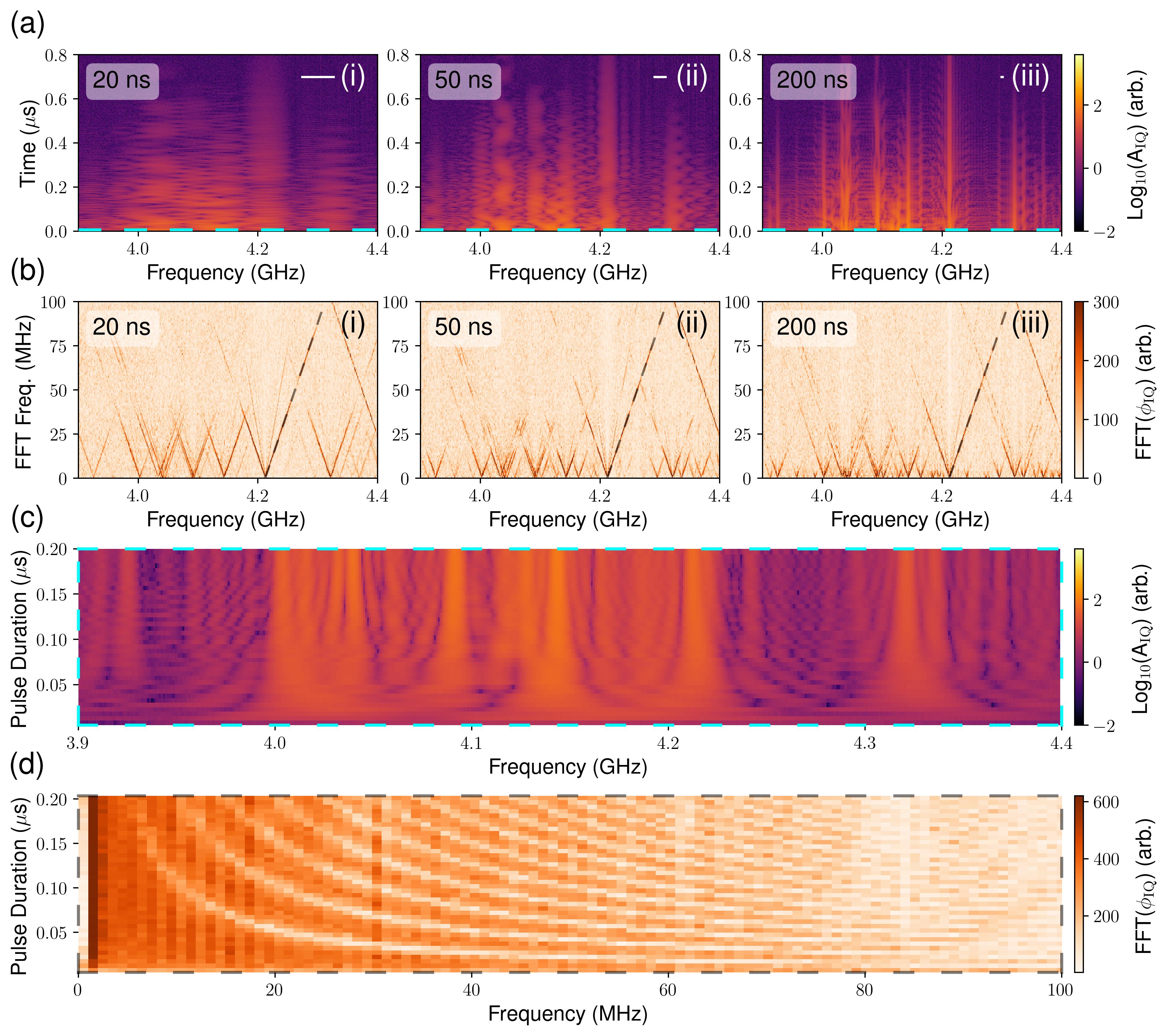}
% \end{centering}
\caption{BCTDS response of a sapphire sample with 2~nm AlO$_x$ deposition under varying pulse durations. (a) Logarithmic amplitude of the transient dielectric response. (b) FFT of the phase of the transient response. For each homodyne frequency, we compute the phase of the demodulated signal and take the FFT of the full post-pulse phase trace. This analysis reveals V-shaped features centered near the bare eigenfrequencies of TLS defect ensembles. A representative slice along the right arm of a prominent V-shaped feature at 4.212~GHz is indicated by the black dashed line. Panels (a) and (b) show three pulse durations: 20~ns (i), 50~ns (ii), and 200~ns (iii). A short horizontal white line indicates the pulse bandwidth $\Delta f$. As the pulse duration increases, the pulse bandwidth decreases, leading to sharper spectral features. (c) Zero-time transient-response slices extracted from the cyan dashed line in (a) at $t=0$, stacked as a function of pulse duration, showing the sharpening of interference fringes as the pulse length increases. (d) Stacked line cuts extracted from the black dashed line in (b), as a function of pulse duration, showing the interference pattern along the V arm as a function of pulse duration. The pulse-duration-dependent sharpening, V-shaped phase-FFT branches, and interference along the V arm are qualitatively reproduced by the driven standard tunneling model in Fig.~\ref{fig:simulation_pulse_width_ringdown_fft}, while the weak-drive analytical limit in Fig.~\ref{fig:analytics_multispin} shows that the basic V-shaped geometry can already emerge from finite-time coherent driving of a disordered, non-interacting TLS defect ensemble.}
\label{fig:pulse_width}
\end{figure*}

The experimental results present the BCTDS response through phase, amplitude, and correlation analyses. Figure~\ref{fig:pulse_width} first tests the phase-sensitive operating principle by varying the pulse duration in a sapphire sample with 2~nm AlO$_x$. This measurement reveals V-shaped phase-FFT features and pulse-duration-dependent interference fringes, which we interpret using the Floquet and finite-time-drive analyses in Appendices~\ref{app:floquet_formalism}, \ref{app:pulse_dependence}, and \ref{app:analytics}. Figure~\ref{fig:sample_compare} then establishes the experimental controls and material dependence of the transient response. Room-temperature measurements of the empty waveguide and solvent-cleaned sapphire show little to no discernible post-pulse emission, while cryogenic measurements reveal long-lived transient features whose amplitude and spectral structure depend on the sample, with stronger responses observed for the sapphire samples with 2~nm AlO$_x$ film and with coated photoresist. The extracted ring-down lifetimes are described in Appendix~\ref{app:exp_fit}, and Appendix~\ref{app:mode_leakage} discusses how the microwave environment can filter or reshape the measured transient field. Figure~\ref{fig:cryo_compare} shows a thermal-cycle comparison experiment that is consistent with the interpretation that the BCTDS response arises primarily from TLS defects instead of spurious cavity modes. Finally, Fig.~\ref{fig:transient_quantification} uses the photoresist dataset from Fig.~\ref{fig:sample_compare}(f) to examine intensity correlations and an effective dissipative response, with the IQ-to-susceptibility connection discussed in Appendix~\ref{app:tls_response}. Together, these measurements show that BCTDS probes the phase structure, amplitude response, and temporal correlations of TLS defects in the sample.

%%%%%%%%%%%%%%%%%%%%%%%%%%%%%%%%%%%%%%%%%%%%%%%%%%%%%%%%%%%%%%%%%%

% \begin{figure*}[htpb]
% \includegraphics[width=0.88\textwidth]{figures/Fig_transient_quantification.png}
% \caption{} 
% \label{fig:transient_quantification}
% \end{figure*}
%%%%%%%%%%%%%%%%%%%%%%%%%%%%%%%%%%%%%%%%%%%%%%%%%%%%%%%%%%%%%%%%%%

\subsection{\label{subsec:different_length_drives} BCTDS Response under Different Pulse Durations}

We begin by focusing on a 2~nm AlO$_x$ film grown on sapphire, which we use as a representative proxy for Josephson-junction oxides in superconducting circuits (more detailed descriptions of material processing and fabrication can be found in Sec.~\ref{subsec:sample_compare}). We study its BCTDS response within the 3.9--4.4 GHz band, where varying the pulse duration allows us to resolve how the transient emission sharpens and develops interference features. In this picture, the finite-length pulse excites a frequency-dependent subset of TLS defects, which subsequently generate transient coherent emission after the drive is turned off. %We therefore use the AlO$_x$ sample as a representative case and study its response within the 3.9--4.4~GHz band, where changing the pulse duration allows us to resolve how the transient emission sharpens and develops interference features. In this picture, the pulse excites a frequency-dependent subset of TLS defects, which subsequently produce transient coherent emissions after the drive is turned off.

Figure~\ref{fig:pulse_width} (a-b) shows the BCTDS response for pulse durations of (i) 20~ns, (ii) 50~ns, and (iii) 200~ns, respectively. In Fig.~\ref{fig:pulse_width}(a), we plot the logarithmic transient amplitude as a function of drive frequency and post-pulse time. For short pulses, the excitation bandwidth is broad, and the transient response is correspondingly diffuse across frequency. As the pulse duration is increased, the response becomes more spectrally selective and develops sharper, longer-lived features. This trend is consistent with the narrower bandwidth of longer pulses and with coherent energy storage and re-emission by subsets of TLS defects, as already suggested by the material-dependent transients in Fig.~\ref{fig:sample_compare}. Because the measured signal is obtained in a broadband waveguide, these features should be interpreted as the response of the coupled TLS defect--microwave environment rather than as a one-to-one map of isolated defect frequencies.

The phase response provides a more direct connection to the mechanism described in Sec.~\ref{sec:op_principle}. During the pulse, the microwave field dresses the TLS defect ensemble and can drive repeated passages through avoided crossings in the instantaneous dressed-state spectrum. In this Landau--Zener--St\"uckelberg picture, amplitudes generated during repeated passages accumulate a phase that depends on detuning and pulse duration. After the drive is switched off, the dressed-state superpositions are projected onto the bare TLS basis and radiate into the waveguide. The homodyne phase then evolves at rates set by the relevant detunings, so the Fourier transform of the post-pulse phase produces V-shaped branches when sweeping the drive frequency. The connection between drive-prepared quasienergy structure, slow post-pulse evolution, and enhanced ring-down is shown more explicitly in Fig.~\ref{fig:floquet_quasienergies} of Appendix~\ref{app:floquet_formalism}.

This behavior is more clearly observed experimentally in Fig.~\ref{fig:pulse_width}(b), where the magnitude of the phase Fourier spectrum reveals V-shaped features over the same frequency range. The vertices of these branches identify TLS defect-sensitive response frequencies, while the arms correspond to increasing phase-evolution rates away from resonance. The V-shaped features exhibit oscillatory structure along the V arms, and this modulation changes with increasing pulse duration as the excitation becomes more spectrally selective. A representative branch is highlighted by the black dashed line. The fact that the same structures appear in the post-pulse phase, rather than only in the driven amplitude, indicates that the measurement is sensitive to phase coherence accumulated during the finite drive.

Figures~\ref{fig:pulse_width}(c) and \ref{fig:pulse_width}(d) isolate the pulse-duration dependence more directly. Figure~\ref{fig:pulse_width}(c) shows zero-time transient slices, taken immediately after the pulse is turned off, stacked as a function of pulse duration. This removes the subsequent ring-down dynamics and emphasizes the state of the emitted field at the end of the drive. The map shows frequency-dependent regions of enhanced and suppressed post-pulse emission, with narrow spectral features becoming more visible for longer pulses. These features indicate that changing the pulse duration modifies both the excitation bandwidth and the phase accumulated by the driven TLS defect ensemble before emission. They also qualitatively resemble the finite-time interference patterns obtained in the weak-drive, non-interacting, analytical limit shown in Fig.~\ref{fig:analytics_multispin}(a) of Appendix~\ref{app:analytics}, supporting the interpretation that the measured modulation reflects coherent phase accumulation across a disordered TLS defect ensemble during the pulse.

Figure~\ref{fig:pulse_width}(d) provides a complementary phase-sensitive view by following the Fourier amplitude along the representative V-shaped branch marked in Fig.~\ref{fig:pulse_width}(b). The resulting map shows oscillatory fringes as a function of pulse duration and Fourier frequency, indicating that the phase-FFT response depends sensitively on the pulse duration. The interacting four-spin simulations in Fig.~\ref{fig:simulation_pulse_width_ringdown_fft}(c) of Appendix~\ref{app:pulse_dependence} reproduce the same qualitative behavior along a selected V arm, where the Fourier amplitude oscillates with pulse duration due to finite-time phase accumulation. Together with the V-shaped spectra in Appendix Figs.~\ref{fig:simulation_pulse_width_ringdown_fft}(b) and \ref{fig:analytics_multispin}(b), these results support the interpretation that the experimental branches and fringes arise from detuning-dependent phase evolution, as described in Sec.~\ref{sec:op_principle}.
Together, Fig.~\ref{fig:pulse_width} connects the material-dependent transient emission of Fig.~\ref{fig:sample_compare} with the phase-FFT operating principle developed in Sec.~\ref{sec:op_principle}. Varying the pulse duration changes both the excitation bandwidth and the phase accumulated during the drive. The observed sharpening of the amplitude response, the V-shaped phase-FFT branches, and the pulse-duration-dependent fringes are consistent with finite microwave pulses preparing drive-dressed TLS defect coherences that are subsequently emitted into the broadband waveguide after the pulse is turned off.

\subsection{\label{subsec:sample_compare} Cryogenic Dielectric Spectroscopy of Different Materials} 

\begin{figure*}[ht!]
\begin{centering}
\includegraphics[width=0.98\textwidth]{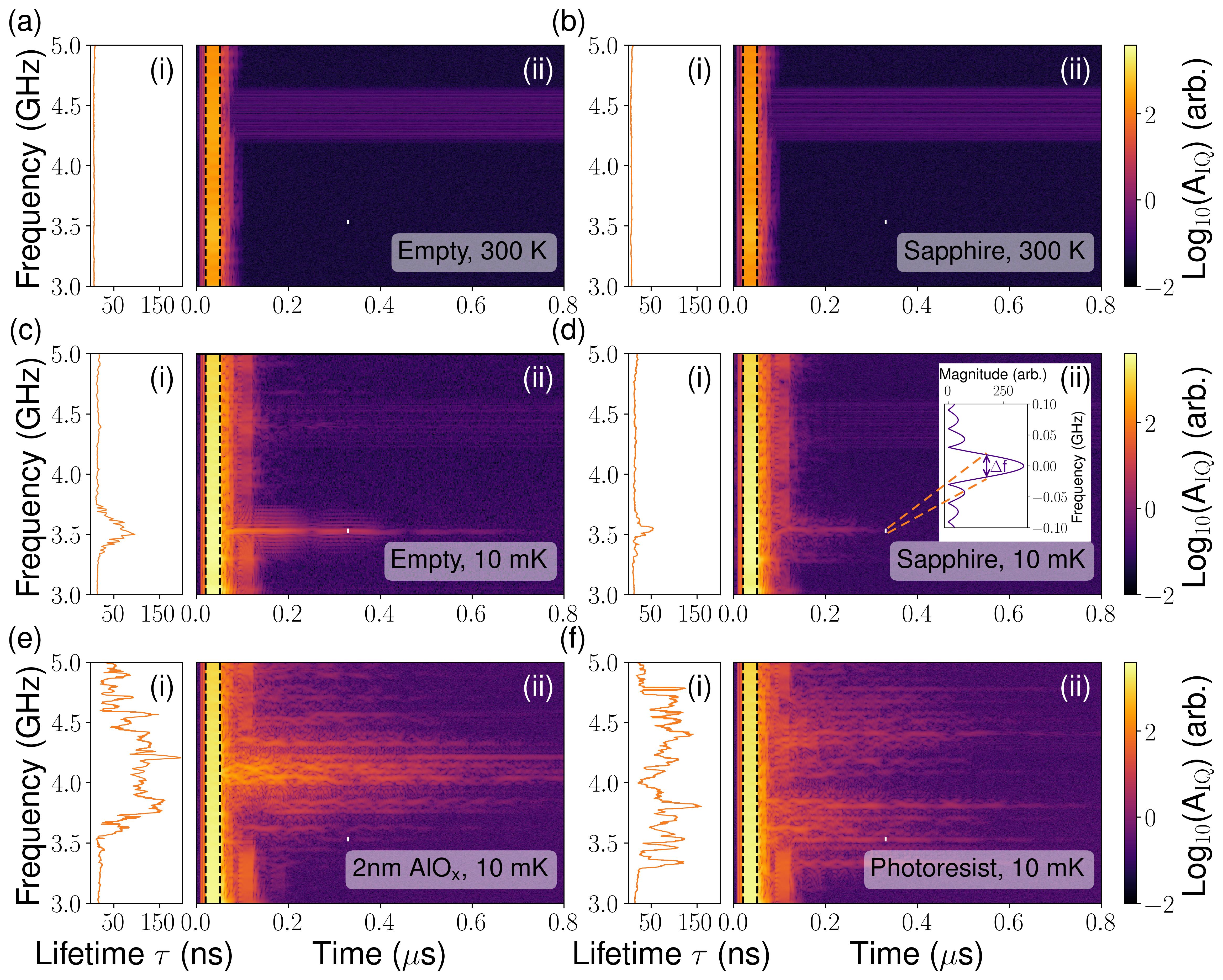}
\end{centering}
\caption{Comparison of transient dielectric responses across different samples. We send a 30 ns pulse (marked by black dashed lines) and readout over a 0.8 $\mu$s window (shown in subpanels (ii) with the color scale clipped to reveal weak transient features). We perform room-temperature control measurements outside the fridge for an empty waveguide (a) and for the waveguide loaded with solvent-cleaned sapphire samples (b), both of which show no detectable transient response. The horizontal band near 4.4 GHz arises from the ADC sampling clock onboard the RFSoC. Measurements (c)--(f) are performed at 10~mK, showing transient dielectric-response features consistent with TLS defect-mediated dynamics. Measurements (c) and (d) feature the same setups as (a) and (b), at 10~mK. (e) The same samples as (b) with a 2~nm AlO$_x$ layer deposited via ALD. (f) The same samples as (b) with spin-coated Shipley 1813 photoresist. The transient response in (e)--(f) shows prominent ring-down features, consistent with an enhanced TLS defect-mediated dielectric response. The HEMT amplifier can briefly saturate during the pulse window, but does not produce long-lived ringdown features (see Appendix~\ref{app:HEMT_sat}). Subpanels (i) show the extracted lifetimes ($\tau$) of the ring-downs, using the exponential fitting procedure described in Appendix~\ref{app:exp_fit}. We include the FFT of the pulse in the inset in (dii) and mark the FWHM of the pulse, $\Delta f$, with a vertical white line, which has approximately the same bandwidth as the prominent spectral features.}
\label{fig:sample_compare}
\end{figure*}

As described in Sec.~\ref{sec:op_principle}, our waveguide enables a modular characterization of the driven and transient dielectric response of ensembles of TLS defects in various samples.  The samples considered in this study started with a 50.8 mm diameter, 432 $\mu$m thick wafer of ultra-high-purity HEMEX sapphire from Crystal Systems, which has been measured to have bulk loss tangents of the order $10^{-8}$ using the dielectric dipper technique \cite{read_precision_2023}. We spin coat the samples with photoresist and dice them into rectangular chips with 27.5 $\times$ 5.5 mm dimensions. We clean the samples with solvents following a standard procedure described in Appendix~\ref{app:SampleProcessing} and mount them in the waveguide.

%% Room temperature.
Figure~\ref{fig:sample_compare} summarizes the BCTDS results obtained for different materials, highlighting the amplitude of the BCTDS signal, in contrast to the phase, to demonstrate the contrast in total signal. We use a short $\sim 30~\mathrm{ns}$ pulse to provide a broadband excitation of the sample and probe its transient response after the drive is removed. This pulse duration is also comparable to the microwave control-pulse timescales commonly used for single-qubit operations in superconducting circuits. Here, we perform room-temperature control experiments on an empty waveguide and on a waveguide with bare sapphire samples that serve as benchmarks, shown in Fig.~\ref{fig:sample_compare}(a) and \ref{fig:sample_compare}(b), respectively. Given that the TLS defects observed in our experiments have frequencies between 3 and 5 GHz, small temperature changes can significantly impact their thermal occupation, a key signature that has been explored in previous work \cite{crowley_disentangling_2023, lisenfeld_measuring_2010}. %Furthermore, increases in temperature cause thermal occupation of phonon modes, which can lead to incoherent shaking of the double-well potential governing the TLS defects, likely broadening or destroying their spectral features. Our room temperature measurements in Fig.~\ref{fig:sample_compare}(aii) and \ref{fig:sample_compare}(bii) confirm this thermal saturation effect.
Furthermore, increasing temperature thermally populates phonon modes, which can induce incoherent fluctuations of the double-well potentials associated with the TLS defects, thereby broadening or suppressing their spectral response. Consistent with this picture, our room-temperature measurements in Fig.~\ref{fig:sample_compare}(aii) and Fig.~\ref{fig:sample_compare}(bii), which do not exhibit the long-lived ringdown features, are indicative of strong thermal saturation effects in the TLS defect ensemble.

%% Sapphire layer.
As motivated in Sec.~\ref{sec:op_principle}, cooling the system suppresses thermal occupation and allows coherent transient features to become visible in the post-pulse response. In the Floquet picture, a finite microwave pulse can populate dressed TLS defect states whose quasienergy differences set the post-pulse phase-evolution rates. Figure~\ref{fig:floquet_quasienergies} of Appendix~\ref{app:floquet_formalism} illustrates this mechanism in a minimal simulation, where enhanced post-pulse tails occur near drive frequencies with small quasienergy differences. This provides a qualitative framework for why spectral features in the driven response can be accompanied by longer-lived ringdowns in the transient response under sufficiently strong driving amplitude, as depicted in Fig.~\ref{fig:simulation_driving_amplitude} in Appendix~\ref{app:amplitude_dependence}. Consistent with this picture, when we cool the same setups in Fig.~\ref{fig:sample_compare}(a) and Fig.~\ref{fig:sample_compare}(b) to 10~mK, we observe apparent narrowing of the spectral dips in the driven dielectric spectroscopy and a corresponding increase in the temporal structure of the transient dielectric response, shown in Fig.~\ref{fig:sample_compare}(c) and Fig.~\ref{fig:sample_compare}(d). These features are consistent with coherent storage and re-emission by ensembles of TLS defects. We further discuss this interpretation using Floquet theory and master-equation simulations in Fig. \ref{fig:floquet_quasienergies} in Appendix~\ref{app:floquet_formalism}. In Fig.~\ref{fig:sample_compare}(c), we observe transient features in the absence of a mounted sample. These features may partly arise from TLS defects in native oxide layers on the waveguide and antenna surfaces. %However, classical electromagnetic effects also likely contribute to the measured response. 
%Although the waveguide is designed to support propagation of the dominant TE$_{10}$ mode over the 3--6~GHz band, a thin sample slot introduces an opening through which microwave fields can couple to the surrounding reflective metal enclosure at the mixing chamber and can lead to formation of cavity-like or harmonic oscillator modes. 
Although the waveguide is designed to support propagation of the dominant TE$_{10}$ mode over the 3--6~GHz band, a thin sample slot introduces an opening through which microwave fields can couple to the surrounding reflective metal enclosure at the mixing chamber and can lead to the formation of cavity-like or harmonic oscillator modes. At cryogenic temperatures, coherently driven TLS defects within the sample can radiatively emit into these modes after the external drive is turned off. Within an input-output description, these modes act as weakly damped resonant channels driven by the collective TLS defect polarization. In this framework, the cavity-like mode does not store TLS defect dynamics directly, but instead evolves as a damped harmonic oscillator whose amplitude is continuously driven by the TLS defect dipole, thereby accumulating a time-integrated and exponentially weighted record of the emitted field. Because these modes can possess linewidths substantially narrower than the broadband waveguide continuum, they can support long-lived microwave ringdown signals extending to microsecond timescales. The measured output, therefore, reflects a delayed, spectrally filtered representation of the TLS defect emission encoded in the dynamics of these weakly damped electromagnetic modes. While our measurements do not directly isolate the role of phonons, TLS defects can couple both to strain fields and microwave electric fields, and TLS--phonon interactions may therefore additionally influence the temporal coherence and relaxation dynamics of the emitted field.
Appendix~\ref{app:mode_leakage} discusses the origin of cavity-like modes in more detail using room-temperature measurements of the waveguide mounted inside the reflective mixing-chamber enclosure. The observed homodyne spectrum should therefore be viewed as the response of a coupled system involving the driven TLS defect ensemble and the surrounding microwave environment. In this regime, classical ringing from the microwave structure can overlap with, filter, or interfere with the TLS defect-induced polarization response. This coupling makes a complete separation of classical and quantum contributions difficult. However, the observed features are strongly sample and temperature-dependent, showing frequency shifts (Appendix~\ref{app:temp_shift}) and suppression of the interference patterns at several hundred millikelvin \cite{MQT}, as well as spectral reconfiguration after thermal cycling (Sec.~\ref{app:cooldown_compare}). Together, these behaviors are consistent with TLS defect-mediated dynamics and motivate BCTDS as a sensitive probe of TLS defect-related transient dielectric response. The signal in Fig.~\ref{fig:sample_compare}(d) likely originates from similar sources but may also include contributions from the solvent-cleaned sapphire samples in the waveguide, where residual surface contamination can result from leftover photoresist, adventitious carbon, or surface and dicing damage. In future work, we intend to remove these baseline contributions by retracting the samples in situ with a piezoelectric positioner, inspired by the dielectric dipper approach \cite{read_precision_2023}. This calibration will enable further exploration of surface treatments such as etching, annealing, and passivation to reduce the measured transient dielectric response of bare sapphire and silicon wafers, which form the basis of all superconducting circuits \cite{place_new_2021, PhysRevX.9.031052, olszewski_low-loss_2025}.

%% Aluminum oxide layer.
Today, most Josephson junctions are fabricated using various versions of double-angle evaporation, where the insulating Josephson junction layer is formed from a native oxide grown on deposited aluminum. However, all regions of the chips that contain aluminum will inadvertently grow such native oxides, and therefore, most superconducting circuits inevitably contain large areas of aluminum oxide. To emulate the native oxide in Josephson junctions and on aluminum metal layers, we grow a thin, 2~nm AlO$_x$ layer in an atomic layer deposition (ALD) tool on a bare sapphire wafer. The results of these BCTDS measurements are shown in Fig.~\ref{fig:sample_compare}(e). Here, we observe pronounced transient features consistent with a TLS defect contribution. 
The transient response following the application and subsequent removal of the monochromatic drive exhibits pronounced collapse-and-revival–like features arising from the collective response of a disordered TLS defect ensemble, in which different defects are driven with varying detunings relative to the applied frequency. As discussed above, we use a relatively strong excitation pulse to drive the collective TLS response. Although this pulse can temporarily saturate the high-electron-mobility transistor (HEMT) amplifier during the pulse window, the readout chain recovers quickly and does not produce the longer-lived ringdown features. This behavior is verified by the bypass control measurement in Appendix~\ref{app:HEMT_sat} (Fig.~\ref{fig:HEMT_bypass}), supporting our attribution of the post-pulse dynamics to the sample response rather than receiver-chain saturation. The coexistence of near-resonant and off-resonant TLS defect contributions can give rise to interference and beating effects that suggest coherent memory effects in the driven dynamics \cite{Lisenfeld2019, Gaikwad2024}. Furthermore, we want to highlight that the deposited 2~nm of AlO$_x$ is approximately $5 \times 10^{-6}$ times the volume of the substrate, but is composed primarily of the same atomic species. These findings highlight the importance of  material chemistry, structural phase, and disorder for the measured TLS defect response and motivate future work on passivating metallic layers \cite{PhysRevLett.134.097001} and developing crystalline insulating layers for Josephson junctions. This work also suggests why tantalum, which hosts a thin, stoichiometric native oxide, is an advantageous material choice for high-coherence superconducting circuits \cite{place_new_2021, bland2025}.

% Resist layer.
Finally, we characterize the dielectric response of 1--4~$\mu$m of Shipley 1813 photoresist on sapphire, as shown in Fig.~\ref{fig:sample_compare}(f). This sample is relevant because most resonators and waveguides are fabricated using photolithography, and residual photoresist can remain on the substrate if it is not fully removed. Photoresists are complex polymer materials and are expected to host a high density of TLS defects~\cite{Quintana2014,Muller2019}; details of the sample preparation are given in Appendix~\ref{app:SampleProcessing}. By measuring the transient response of the resist layer with BCTDS, we can assess how such residues may contribute to dielectric loss and inform fabrication and device-design optimization~\cite{megrant_scaling_2025}. In Fig.~\ref{fig:sample_compare}(f), the photoresist-coated sapphire shows an enhanced post-pulse transient response, consistent with a higher density of TLS defect-mediated dielectric activity in the resist layer.

More broadly, the technique introduced here enables longitudinal studies in which the same sample can be characterized at different stages of the fabrication process. Such measurements could help identify processing steps that increase or suppress TLS defect-related transient response, providing a route toward active mitigation strategies for future superconducting-qubit devices. Across the datasets in Fig.~\ref{fig:sample_compare}, the observed transient emission has characteristic lifetimes on the order of $100$~ns. This is shorter than the $\sim 1~\mu$s dielectric-echo lifetimes reported in recent JTWPA experiments~\cite{boselli2025,delattre_quantitative_2025}. In our broadband waveguide geometry, the larger electromagnetic density of states and the collective nature of the measurement may reduce the observed lifetimes through Purcell-enhanced decay and stimulated emission. Although these lifetimes are short compared with state-of-the-art superconducting-qubit coherence times, the corresponding transient response could still be relevant in larger quantum circuits where many gates are applied sequentially. Such effects may contribute to the gap between system-level fidelities and estimates based only on independently measured single- and two-qubit gate fidelities from randomized benchmarking~\cite{PhysRevA.75.022314}.

\subsection{Thermal Cycle Dependence}
\label{app:cooldown_compare}

%%%%%%%%%%%%%%%%%%%%%%%%%%%%%%%%%%%%%%%%%%%%%%%%%%%%%%%%%%%%%%%%%%

\begin{figure*}[ht!]
\begin{centering}
\includegraphics[width=0.98\textwidth]{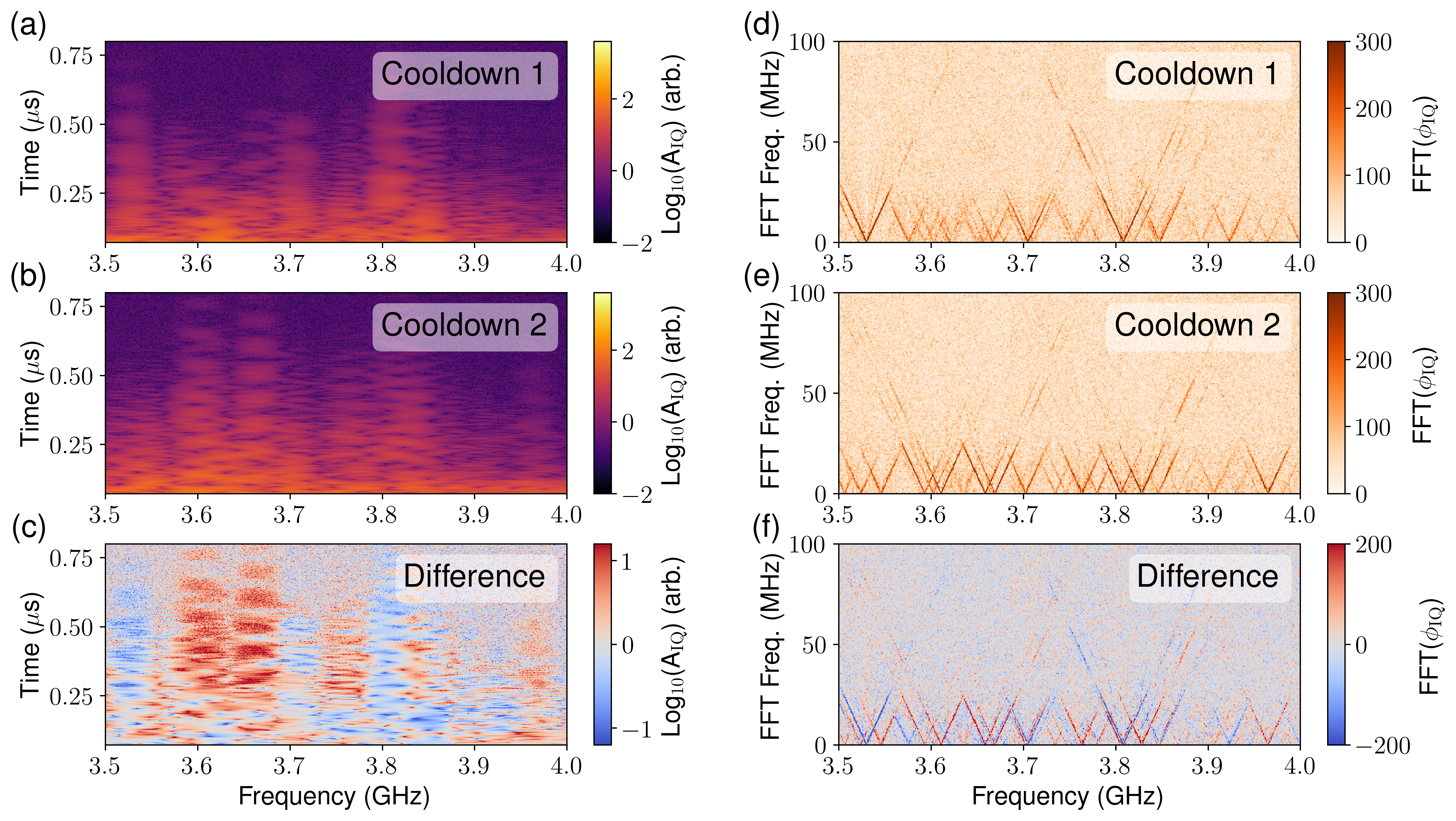}
\end{centering}
\caption{Cooldown-to-cooldown comparison of the cryogenic transient dielectric response of Shipley 1813 photoresist on sapphire, zoomed in to the 3.5--4.0~GHz range to resolve finer spectral features.
(a) Logarithmic
amplitude response of cooldown 1, using the same dataset as Fig.~\ref{fig:sample_compare}(f).
(b) Logarithmic
amplitude response of cooldown 2, measured on the same sample after warming to 300~K and cooling back to base temperature.
(c) Difference between the two cooldowns, computed as the difference between the $\log_{10}$ transient response magnitudes, revealing a strong change in the spectral fingerprint of the transient response.
(d) FFT of phase transient response for cooldown 1.
(e) FFT of phase transient response for cooldown 2.
(f) Difference between the FFT of phase responses in (d) and (e), clearly showing changes in the resonance structure between the two cooldowns.
Importantly, the striking differences highlighted in (c) and (e), where the measurement setup was unmodified, point to the conclusion that the BCTDS response likely arises from TLS defects that reconfigure under thermal cycling rather than spurious cavity modes. }
\label{fig:cryo_compare}
\end{figure*}

%%%%%%%%%%%%%%%%%%%%%%%%%%%%%%%%%%%%%%%%%%%%%%%%%%%%%%%%%%%%%%%%%%

TLS defects are known to shift in frequency after thermal cycling \cite{PhysRevLett.105.177001} likely due to rearrangements in the local environment. To corroborate this signature, we perform BCTDS 15 days apart on the same sample, without changing the measurement setup, subject to a thermal cycle to 300 K in Fig.~\ref{fig:cryo_compare}. As expected, we see a significantly different transient dielectric response in Fig.~\ref{fig:cryo_compare}(a) and \ref{fig:cryo_compare}(b), highlighted by the difference plot in Fig.~\ref{fig:cryo_compare}(c). To resolve the frequency components more clearly, we plot the FFT of phase responses for cooldowns 1 and 2 in Figs.~\ref{fig:cryo_compare}(d) and \ref{fig:cryo_compare}(e), respectively, and their difference in Fig.~\ref{fig:cryo_compare}(f). The FFT of phase difference plot Fig.~\ref{fig:cryo_compare}(f) clearly reveals a reconfiguration of the resonance structure through changes in the positions and prominence of the V-shaped features. These measurements provide further support that the response originates from TLS defects instead of cavity modes.

\subsection{\label{sec:level3} Further Investigation of Cryogenic Transient Dielectric Spectroscopy Results}

Among the samples measured in Fig.~\ref{fig:sample_compare}, the photoresist-coated sapphire sample exhibits a pronounced post-pulse response. We therefore use this dataset to analyze temporal correlations in the emitted field. The previous sections showed that finite microwave pulses generate spectrally structured emission, including V-shaped phase-FFT features and pulse-duration-dependent interference patterns. Here, we ask whether this emitted field also retains measurable correlations after the drive is turned off. This extends the analysis from spectral features to time-domain correlations in the transient dielectric response.

%%%%%%%%%%%%%%%%%%%%%%%%%%%%%%%%%%%%%%%%%%%%%%%%%%%%%%%%%%%%%%%%%%
\begin{figure*}[htpb]
% \begin{centering}
\includegraphics[width=0.88\textwidth]{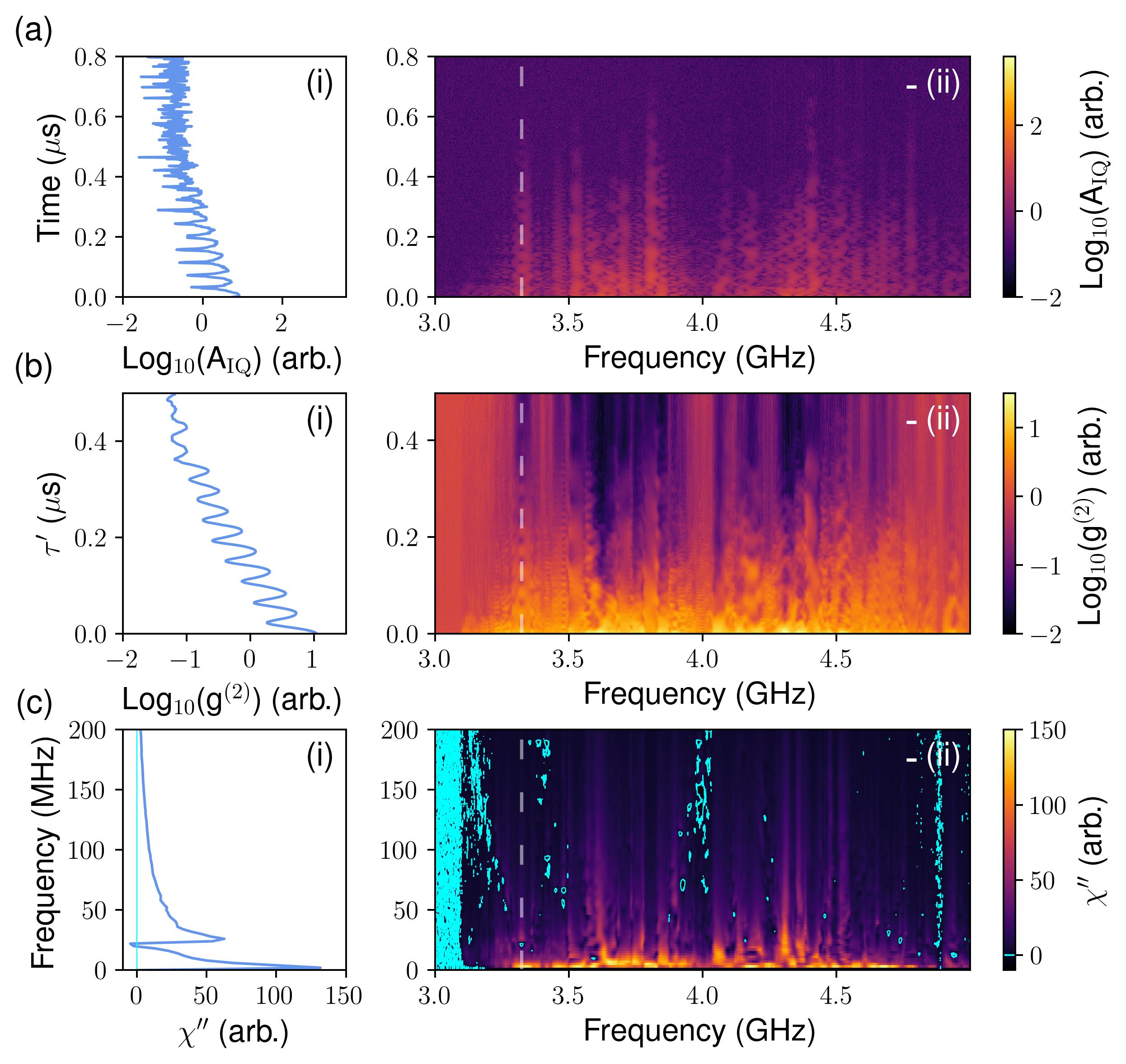}
% \end{centering}
\caption{Transient dielectric response of Shipley 1813 photoresist on sapphire. We use the same data as Fig.~\ref{fig:sample_compare}(f), and panels (a)--(c) are different analyses of the same measured dataset. (aii) Logarithmic magnitude of the transient dielectric response. (bii) Logarithm of the two-time correlation function $g^{(2)}$ map of the transient spectrum, defined by Eq.\,\ref{eq:g2}. (cii) $\chi''$ computed from $g^{(2)}$ using Eq.\,\ref{eq:chi_imag_IQ}. We highlight transitions from positive to negative $\chi''$ with the cyan contour in (cii). Subpanels (i) show linecuts at $\omega/2\pi = 3.322$~GHz, marked by the dashed line in (ii), highlighting temporal structure in the ring-down: collapse-and-revival-like modulations in (a), non-exponential decay, bunching, and oscillations in $g^{(2)}$ in (b), and a non-monotonic effective $\chi''$ with sign changes in (c). These features are suggestive of memory effects in the transient response, but they should not be interpreted as a complete measure of non-Markovianity. We also note that $\chi''$ oscillates around zero near $\omega/2\pi = 3.0$~GHz primarily due to weak signals and does not by itself indicate memory effects. A short horizontal white line indicates the pulse bandwidth $\Delta f$.} 
\label{fig:transient_quantification}
\end{figure*}
%%%%%%%%%%%%%%%%%%%%%%%%%%%%%%%%%%%%%%%%%%%%%%%%%%%%%%%%%%%%%%%%%%

Figure~\ref{fig:transient_quantification}(a-ii) shows the post-pulse homodyne amplitude from the photoresist-coated sapphire sample, plotted as $\log_{10}(A_{IQ})$. This is the same transient response shown in Fig.~\ref{fig:sample_compare}(f), now used as the starting point for the correlation analysis. The signal decays over the measured post-pulse window, but it also contains frequency-localized features that remain visible beyond the initial emission. The representative line cut at $\omega/2\pi = 3.322$~GHz, shown in Fig.~\ref{fig:transient_quantification}(a-i), makes the time-domain structure clear. The ring-down contains oscillatory modulations and partial revivals on top of the overall decay, rather than a smooth single-exponential relaxation. As in Sec.~\ref{subsec:different_length_drives}, we interpret the measured signal as the response of the coupled TLS defect--microwave environment, since the waveguide can mediate, filter, and overlap with the emitted transient field.

To quantify the temporal structure of this emission, we compute the normalized two-time intensity correlation function
\begin{equation}
g^{(2)}(\tau') =
\frac{
\langle \mathcal{I}(t)\mathcal{I}(t+\tau') \rangle
}{
\langle \mathcal{I}(t) \rangle^2
},
\label{eq:g2}
\end{equation}
where $\mathcal{I}(t)\propto A_{IQ}^2(t)$ is the measured homodyne intensity , $\tau'$ is the correlation time delay variable and $\langle \cdots \rangle_t$ denotes an average over the post-pulse time window. Figure~\ref{fig:transient_quantification}(b-ii) shows $\log_{10}[g^{(2)}(\tau')]$ as a function of drive frequency and delay time. The correlation map contains frequency-dependent structures that extend over hundreds of nanoseconds. The line cut at $\omega/2\pi = 3.322$~GHz in Fig.~\ref{fig:transient_quantification}(b-i) shows a non-exponential decay with oscillations, consistent with the modulations observed directly in the ring-down in Fig.~\ref{fig:transient_quantification}(a-i). These features suggest that the transient response is not fully described by an immediate, memoryless relaxation process. Rather, the emitted field retains correlations associated with the driven evolution prepared during the finite pulse.

We then use the same correlation function to estimate an effective dissipative response through Eq.~\ref{eq:chi_imag_IQ}. Figure~\ref{fig:transient_quantification}(c-ii) shows the resulting $\chi''$ as a function of drive frequency and Fourier frequency, with cyan contours marking sign changes. In equilibrium linear response, $\chi''$ corresponds to the absorptive component of the susceptibility. In the present experiment, the system is driven by a finite microwave pulse and measured in the post-pulse regime, so $\chi''$ should be interpreted as an effective measure of the correlated transient dielectric response rather than as a strictly equilibrium susceptibility. The line cut in Fig.~\ref{fig:transient_quantification}(c-i) shows a structured, non-monotonic response with sign changes, consistent with the oscillatory correlations observed in $g^{(2)}(\tau')$. We do not interpret small sign changes near the low-frequency edge of the measurement band, where the signal is weak, as evidence of memory effects by themselves.

Taken together, Fig.~\ref{fig:transient_quantification} shows that the photoresist-coated sapphire sample produces not only a strong transient amplitude but also a correlated post-pulse response. The oscillatory ring-down, the delayed correlations in $g^{(2)}(\tau')$, and the structured effective $\chi''$ are consistent with coherent storage and re-emission in the driven TLS defect ensemble, modified by the broadband microwave environment. These results show that BCTDS captures complementary aspects of the same transient dielectric response. Figure~\ref{fig:sample_compare} identifies the sample-dependent post-pulse emission, Fig.~\ref{fig:pulse_width} shows how this emission changes with pulse duration, and Fig.~\ref{fig:transient_quantification} shows that the emitted field retains temporal correlations during the ring-down.

\section{Conclusions and Outlook}

We have introduced BCTDS as a modular technique for probing the driven and post-pulse dielectric response of materials at cryogenic temperatures. The method uses a broadband 3D waveguide and homodyne detection to measure the transient field emitted after a finite microwave pulse. This geometry allows samples to be mounted and characterized directly, without fabricating narrow-band resonators or qubits, while still accessing frequency- and time-dependent features of the low-temperature dielectric response.

Our measurements show that the transient response depends strongly on temperature and material processing. Room-temperature measurements of the empty waveguide and solvent-cleaned sapphire show little post-pulse structure, while cryogenic measurements reveal clear transient emission. The response becomes more pronounced for samples containing 2~nm AlO$_x$ and Shipley 1813 photoresist on sapphire, consistent with an enhanced low-temperature dielectric response in these materials. Since the measured homodyne signal can also include contributions from the surrounding microwave environment, we interpret the observed spectra as the response of a coupled TLS defect--microwave system rather than as isolated TLS defect emission alone.

By varying the pulse duration, we showed that finite microwave pulses modify both the spectral selectivity and the phase structure of the transient response. Longer pulses sharpen the amplitude response and reveal clearer V-shaped features in the phase FFT. These V-shaped branches are consistent with detuning-dependent phase evolution of driven TLS defect components, as described by the operating principle and supported by the analytical and numerical calculations in the Appendix. The pulse-duration-dependent fringes further indicate that the emitted field retains information about the phase accumulated during the drive.

We also analyzed temporal correlations in the transient emission from photoresist-coated sapphire. The measured ring-down contains oscillatory modulations and partial revivals, while the two-time correlation function shows delayed correlations over hundreds of nanoseconds. The effective $\chi''$ extracted from these correlations is structured and non-monotonic, consistent with a transient dielectric response beyond a simple exponential relaxation process. These features are suggestive of memory effects in the driven ensemble response, although they do not by themselves constitute a complete measure of non-Markovianity. If similar correlated transient effects are present in quantum processors, they could be relevant for quantum error correction and for understanding errors beyond simple Markovian noise models~\cite{PhysRevA.71.012336,Shor1996FaultTolerant,Gaikwad2024}.

Our results support a consistent picture in which finite microwave pulses prepare non-equilibrium dynamics in an ensemble of TLS defects, and the subsequent emission into a broadband waveguide carries spectral and temporal information about that response. The observed transient structure likely reflects the combined influence of driven TLS defect dynamics, possible TLS--TLS interactions, and the surrounding microwave environment. In this sense, BCTDS does not replace resonator, qubit, or dielectric-dipper measurements. Instead, it opens a complementary measurement regime: broadband, strongly driven, time-resolved spectroscopy of TLS-hosting materials before full device fabrication. Although absolute extraction of microscopic defect densities will require calibrated field participation and background subtraction, the present implementation already provides a non-invasive way to fingerprint non-equilibrium ensemble dynamics and their sensitivity to material processing and electromagnetic environment.

Several extensions follow naturally from this work. An orthogonal SMA port could be added to rotate the microwave polarization and repeat the same spectroscopy under different drive strengths and field orientations. Polarization-dependent measurements would help constrain the dipole orientations of the participating defects and provide a more complete picture of their frequency distribution. More broadly, the modularity of BCTDS makes it well suited for longitudinal studies in which the same sample is measured across processing steps such as cleaning, oxide growth, resist coating, etching, annealing, and passivation. This capability could also be extended to measuring samples with patterned films and, with additional calibration, to superconducting devices. In such devices, BCTDS would not directly determine a qubit lifetime, but could help identify off-resonant TLS ensembles, dielectric echoes, or correlated transient responses that may contribute to device loss and non-Markovian noise. Such measurements could help identify which fabrication steps increase or suppress transient dielectric response, providing direct feedback for the development of lower-loss materials and interfaces.

Although this work focuses on TLS defects relevant to superconducting quantum circuits, the technique is more general. Broadband transient spectroscopy may also be useful for probing other microwave-active defects and dielectric excitations, including vacancies and dopants in silicon or nitrogen-vacancy-related systems in diamond, without requiring optical interrogation~\cite{Zhang2024}. In this sense, BCTDS provides a flexible platform for studying how microscopic defect ensembles respond under strong microwave driving. In particular, the measured response may be interpreted as a drive-induced transition from localized to delocalized dynamics in interacting TLS defect ensembles, consistent with the accompanying change in Floquet quasienergy statistics from Poisson-like to circular-ensemble behavior observed in simulations (Fig.~\ref{fig:level_statistics} of Appendix~\ref{app:floquet_formalism}).

\section*{Data Availability Statement}

The data that support the findings of this study are available
from the corresponding author upon reasonable request.

\begin{acknowledgments}
Bert Harrop graciously diced the devices at Princeton. Chloe Buschmann helped review and polish the manuscript. We also acknowledge stimulating conversations with Alexander Carney, Tian Xia, Kater Murch, Miles Blencowe, Geoffroy Hautier, Chandrasekhar Ramanathan, Valla Fatemi, Ioan Pop, Mathieu Fechant, and Lorenza Viola. Startup funds from the Thayer School of Engineering at Dartmouth primarily supported this work. While the scientific focus of this work is quite distinct from existing funding, M.F. would like to express gratitude for the support of science within his group through the DARPA Young Faculty Award No.\, D23AP00192 and NSF PHY-2412555, and hopes that projects like this continue to be made possible by grants from the US federal government. The views and conclusions in this manuscript are those of the authors. They should not be interpreted as representing the official policies, expressed or implied, of DARPA, the NSF,
or the US Government. The US Government is authorized to reproduce and distribute reprints for Government purposes, notwithstanding any copyright notation
herein.
\end{acknowledgments}

\section*{\label{sec:author-contribution} Author Contributions}
M.F. and S.B. contributed to the conceptual development of the study, with M.F. guiding the experimental implementation stack and S.B. guiding the theoretical and numerical simulation stack.
M.F. proposed the broadband waveguide approach and guided the experimental work.
S.B. guided theoretical and numerical simulation work. 
Q.W. performed the experiments, made the figures, and developed the data analysis 
pipeline and performed all necessary sanity checks for this work. 
J.S.S.G and Q.W worked on experimental aspects of microwave engineering and built an app for the lab to estimate attenuation levels needed and the corresponding E.M environment temperature. 
S.M.G., S. A., and J.F. designed and fabricated the waveguide. Q.W, S.A.A, and W.S performed AlO$_x$ deposition. J.S.S.G., R.L., and S.B. provided theoretical support. J.S.S.G., S.B., and M.F. provided context for the work.  J.S.S.G and SB developed theoretical insights using an interacting-driven TLS model to validate the observed patterns. J.S.S.G built the Floquet simulation pipeline and provided software support.
All authors jointly validated the results and participated in the writing of the manuscript.

 \section*{Competing Interests}
The authors declare no competing interests.

\section*{\label{sec:additional-information} Additional information}

\subsection*{Correspondence and request for materials}

Correspondence and material requests should be directed to Mattias Fitzpatrick (mattias.w.fitzpatrick@dartmouth.edu). Data supporting the findings of this study are available from the corresponding author upon reasonable request.

\appendix

%%%%%%%%%%%%%%%%%%%%%%%%%%%%%%%%%%%%%%%%%%%%%%%%%%%%%%%%%%%%%%%%%%
\begin{figure}[ht]
\begin{centering}
\includegraphics[width=0.46\textwidth]{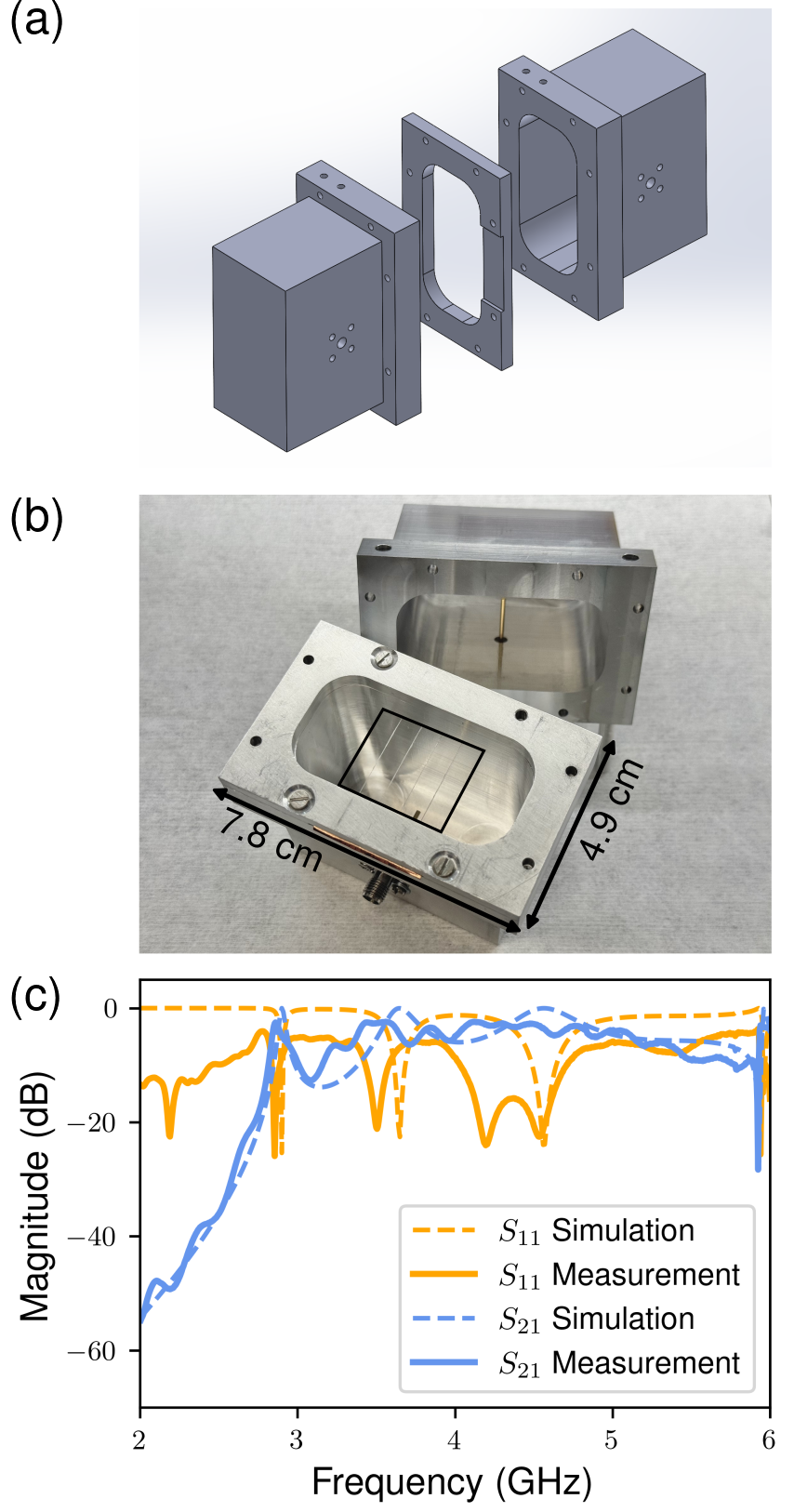}
\end{centering}
\caption{\label{fig:waveguide_transmission} Broadband waveguide design.
(a) Exploded view of waveguide assembly. 
(b) Photograph of upper and lower WR-229 to SMA adapter components and clamp with sapphire samples (marked by the black box). These components are assembled to create the closed 3D aluminum waveguide used in this work.
(c) HFSS simulated and measured transmission ($S_{21}$) and reflection ($S_{11}$) spectra of the waveguide containing samples at room temperature, demonstrating a waveguide cutoff of around 3 GHz and broadband transmission from 3-6 GHz.}
\label{fig:waveguide}
\end{figure}
%%%%%%%%%%%%%%%%%%%%%%%%%%%%%%%%%%%%%%%%%%%%%%%%%%%%%%%%%%%%%%%%%%

\section{Details of the Experimental Setup}
\label{app:WaveguideDesignandFab}

%%%%%%%%%%%%%%%%%%%%%%%%%%%%%%%%%%%%%%%%%%%%%%%%%%%%%%%%%%%%%%%%%%
\begin{figure*}[htpb]
% \begin{centering}
\includegraphics[width=0.88\textwidth]
{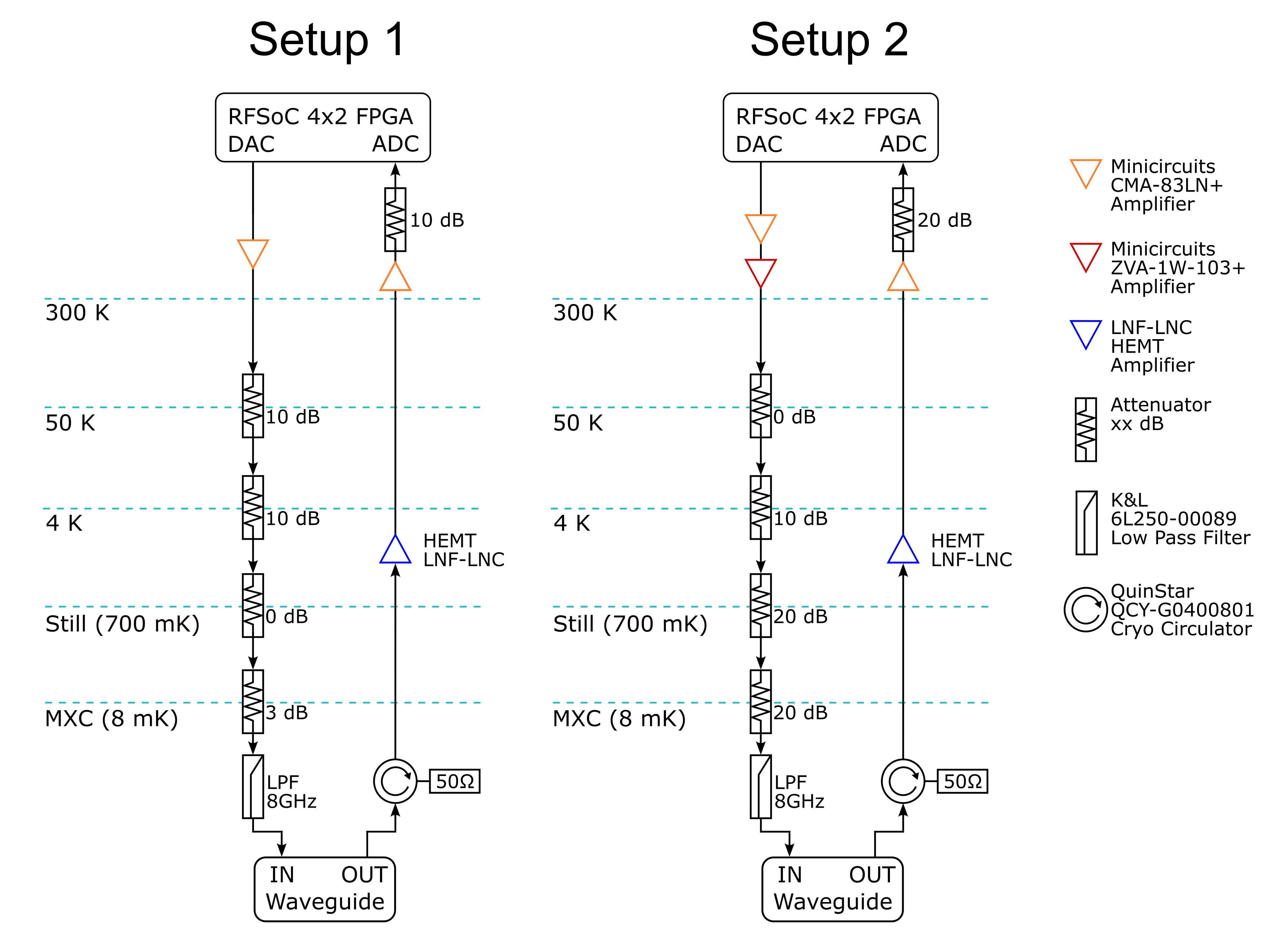}
\caption{Measurement setups for transient dielectric spectroscopy. We use setup 1 for most of the experiments presented in this work, except Fig.\ref{fig:room_cryo_compare}, Fig.~\ref{fig:temperature_shift}, and Fig.~\ref{fig:HEMT_bypass}, which uses setup 2. The CMA-83LN+ amplifier (orange) and the ZVA-1W-103+ amplifier (red) provide approximately $20~\mathrm{dB}$ and $40~\mathrm{dB}$ of gain, respectively. We use a cryogenic circulator (QCY-G0400801) and a HEMT amplifier (LNF-LNC4\_8C), which provides approximately $40~\mathrm{dB}$ of gain, on the readout line to improve readout quality. In setup 1, the RFSoC DAC output power is approximately $-15~\mathrm{dBm}$, and the drive line contains $23~\mathrm{dB}$ of attenuation. In setup 2, the RFSoC DAC output power is approximately $-30~\mathrm{dBm}$, and the drive line contains $50~\mathrm{dB}$ of attenuation. After amplification and an estimated $15~\mathrm{dB}$ of cable loss, both setups deliver approximately $-35~\mathrm{dBm}$ to the waveguide, providing a strong drive for TLS defects.}
\label{fig:fridge}
\end{figure*}
%%%%%%%%%%%%%%%%%%%%%%%%%%%%%%%%%%%%%%%%%%%%%%%%%%%%%%%%%%%%%%%%%

\subsection{Waveguide Design and Fabrication}\label{subsec:waveguide_design}

The fundamental building block of our waveguides is an adapter that provides a 50 $\Omega$ impedance-matched conversion between an SMA coaxial cable and a WR-229 3D rectangular waveguide,  shown in Fig.~\ref{fig:waveguide}(a). To understand the basic features of the waveguide, we solve the wave equation for a rectangular prism, with boundary conditions that the electric field goes to zero at the walls of the waveguide. After performing separation of variables, we can identify a set of allowed eigenmodes for the cross section perpendicular to the direction of propagation, with wavenumbers given by
\begin{equation}
\label{eq:k_c}
    k_c = \sqrt{\left( \frac{m\pi}{a} \right)^2 + \left( \frac{n\pi}{b} \right)^2},
\end{equation}
where $a$ and $b$ are the width and height of the rectangular cross-section of the waveguide, respectively. The constants $m$ and $n$ are integers whose combinations identify different eigenmodes of the cross-section. This means that the propagation constant of the transmitted wave can be expressed as 
\begin{equation}
\label{eq:beta}
    \beta=\sqrt{k^{2}-k_{c}^{2}},
\end{equation}
where  $k$  is the wavenumber. Significantly, Eq.\,\ref{eq:beta} predicts that when $k^2<k_c^2$, $\beta$ becomes imaginary, and the solutions become decaying exponentials along the waveguide in a domain known as the waveguide's cutoff \cite{pozar}. Therefore, we can describe the frequencies of the allowed modes of the waveguide using 
\begin{equation}
\label{eq:f_c_mn}
    f_{c_{mn}} = \frac{k_{c} c}{2\pi},
\end{equation}
where $c$ is the speed of light. From Eq.\,\ref{eq:f_c_mn}, we then identify the cutoff frequency, which is the lowest frequency $\text{TE}_{10}$ mode, given by
\begin{equation}
f_{c_{10}} = \frac{c}{2a},
\label{eq:cutoff}
\end{equation} assuming that $a\geq b$. 

The cutoff frequency formula provided in Eq.\,\ref{eq:cutoff} is idealized and assumes a perfect rectangular waveguide, which is challenging to manufacture in practice due to the sharp inner corners. Therefore, we utilize Ansys HFSS to consider the design of the actual waveguide in this work. The drive and readout ports are coaxial SMA cables that transition into small antennas integrated into the waveguide. This coupling scheme smoothly transforms the coaxial mode of the SMA cables to a close approximation the $\text{TE}_{10}$ mode of an ideal rectangular waveguide, producing a polarized electric field that will ultimately couple to the TLS defects in the sample. The waveguide is machined from  6061 aluminum using a HAAS Super Mini Mill. We improve the surface finish, especially near the rounded corners, using steel wire wool and polishing stones, and clean the final parts with isopropanol (IPA) to remove residues. The full device consists of two SMA-to-WR-229 3D rectangular waveguide adapters with an interchangeable sample clamp mounted between them, allowing different sample geometries and sizes to be tested. A photograph of the finished waveguide assembly with mounted samples is shown in Fig.~\ref{fig:waveguide}(b), and the measured S-parameters are shown in Fig.~\ref{fig:waveguide}(c).

\subsection{Sample Processing}
\label{app:SampleProcessing}

To demonstrate the modularity of our platform, we consider three types of samples, the preparation of which is explained in this section. 

We start with high-grade crystalline 0.5 mm-thick sapphire wafers from Crystal Systems, which are then diced into 27.5 x 5.5 mm rectangular strips, the standard substrate size for hosting 3D transmons. The samples are initially covered in resist AZ 1518 to protect them from debris during dicing (likely another source of TLS defects). We soak the chips in acetone at 70~$^\circ$C overnight and sonicate them in acetone before transferring them to IPA and drying with compressed nitrogen gas.

To study the effects of photoresist on the TLS defect density, we spin-coat Shipley 1813 with thicknesses ranging from 1 to 4 $\mu$m. We also study the oxide layers (typically a few nanometers in thickness), another candidate source of TLS defects. Instead of evaporated aluminum, which would give both an aluminum and an oxide layer, we deposited an approximately 2~nm AlO$_x$ thin film using thermal ALD in an Anric AT410 system. The deposition followed a standard thermal ALD process using trimethylaluminum as the aluminum precursor and H$_2$O as the oxidant. The film was grown using 20 ALD cycles at 80~$^\circ$C. While exposure to water is typically avoided in superconducting qubit fabrication, Here, it is intentionally used as a conventional thermal ALD oxidant to produce an amorphous, disordered AlO$_x$ layer on the clean, high-purity sapphire substrate, likely with residual hydroxyl species incorporated in the film.

\subsection{Measurement Setup} 
\label{app:measurement_setup}

\begin{table}[ht]
    \centering
    \caption{Summary of measured samples and measurement setups.}
    \label{tab:sample_measurement_summary}
    \begin{tabular}{llll}
        \hline
        Figure & \makecell[l]{Measured\\sample} & \makecell[l]{Number of\\sample strips} & \makecell[l]{Measurement\\circuitry} \\
        \hline
        Fig.~\ref{fig:sample_compare}(a,c) & Empty & 0 & Setup 1 \\
        \hline
        Fig.~\ref{fig:sample_compare}(b,d) & Clean sapphire & 5 & Setup 1 \\
        \hline
        \makecell[l]{Fig.~\ref{fig:pulse_width}\\Fig.~\ref{fig:sample_compare}(e)} & \makecell[l]{Sapphire with\\ALD AlO$_x$} & 6 & Setup 1 \\
        \hline
        \makecell[l]{Fig.~\ref{fig:e_field_calibration}} & \makecell[l]{Sapphire with\\ALD AlO$_x$} & 4 & Setup 1 \\
        \hline
        \makecell[l]{Fig.~\ref{fig:sample_compare}(f)\\Fig.~\ref{fig:cryo_compare}\\Fig.~\ref{fig:transient_quantification}\\Fig.~\ref{fig:transmission_reflection}\\Fig.~\ref{fig:exp_fit}} & \makecell[l]{Sapphire with\\Shipley photoresist} & 5 & Setup 1 \\
        \hline
        \makecell[l]{Fig.~\ref{fig:room_cryo_compare}\\Fig.~\ref{fig:temperature_shift}} & \makecell[l]{Sapphire with\\Shipley photoresist} & 1 & Setup 2 \\
        \hline
        \makecell[l]{Fig.~\ref{fig:HEMT_bypass}} & \makecell[l]{Bypassed} & 0 & Setup 2 \\
        \hline
    \end{tabular}
    \label{table:setup}
\end{table}
This work utilizes an AMD radio-frequency system on a chip (RFSoC) board, which provides a fast sampling rate (9.85 GSPS for DAC and 5 GSPS for ADC) and direct synthesis and readout of microwave pulses without needing up- and down-conversion. To control the RFSoC, we utilize the Quantum Instrumentation Control Kit, a Python package developed by Fermi National Accelerator Laboratory \cite{stefanazzi_qick_2022}. 

The pulses generated from the RFSoC pass through a room-temperature amplifier (Minicircuits CMA-83LN+) and a series of attenuators at various temperature stages of a Bluefors LD400 dilution refrigerator. The signal reaches the waveguide input port at base temperature (typically $<$ 10~mK) and interacts with the sample. The output signal from the waveguide is collected using a readout SMA pin that then passes through a cryogenic circulator, HEMT amplifier, and two room-temperature amplifiers (Minicircuits CMA-83LN+), before returning to the RFSoC. A detailed wiring diagram is shown in Fig.~\ref{fig:fridge}, including the two measurement configurations used in this work. Both setups deliver approximately -35 dBm drive to the waveguide. The corresponding sample mounts and measurement setups are summarized in Table~\ref{table:setup}.

\begin{figure}[htpb]
\includegraphics[width=0.48\textwidth]
{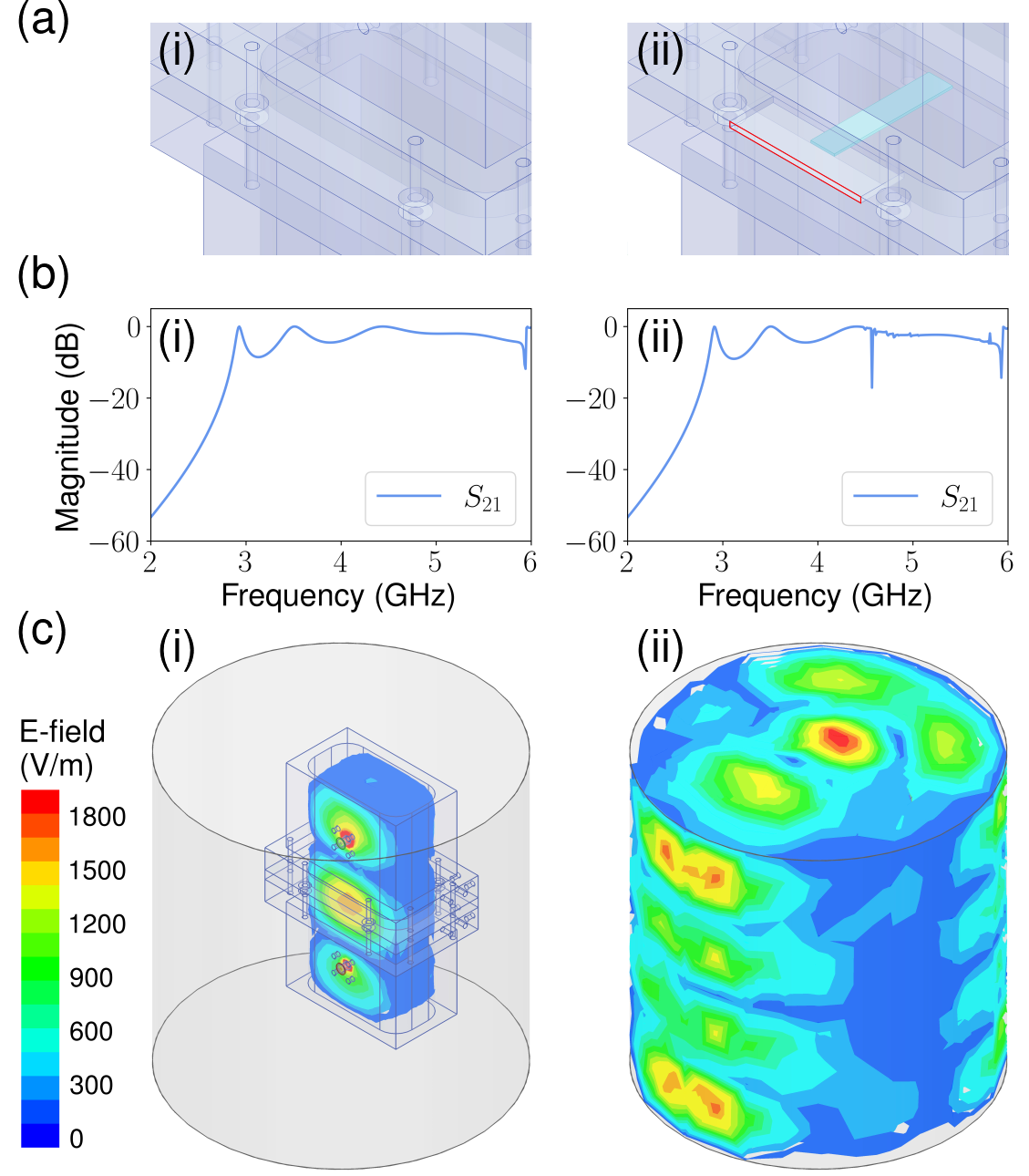}
\caption{Mode leakage through the sample-slot opening.
(a) Zoomed-in waveguide schematics for a waveguide without a sample slot (i) and with a sample slot and mounted sample (ii).
(b) HFSS simulation of the $S_{21}$ transmission for the two geometries shown in (a). Additional transmission dips appear for the slotted waveguide, consistent with coupling to cavity-like modes outside the waveguide.
(c) HFSS simulations of the electric field for a waveguide placed inside a cylindrical conducting can. The field is shown at 4.5~GHz and zero phase for the closed (i) and slotted (ii) waveguides. In the slotted case, microwave fields leak out of the waveguide and couple to the surrounding enclosure, producing cavity-like ringing within the conducting outer can. The simulated cylinder is smaller than the actual mixing-chamber can and omits finer internal structures, such as copper racks, coax cables, and other hardware, serving as a qualitative demonstration of mode leakage and cavity-mode formation.
}
\label{fig:mode_leakage}
\end{figure}

\begin{figure*}[htbp!]
\includegraphics[width=0.88\textwidth]{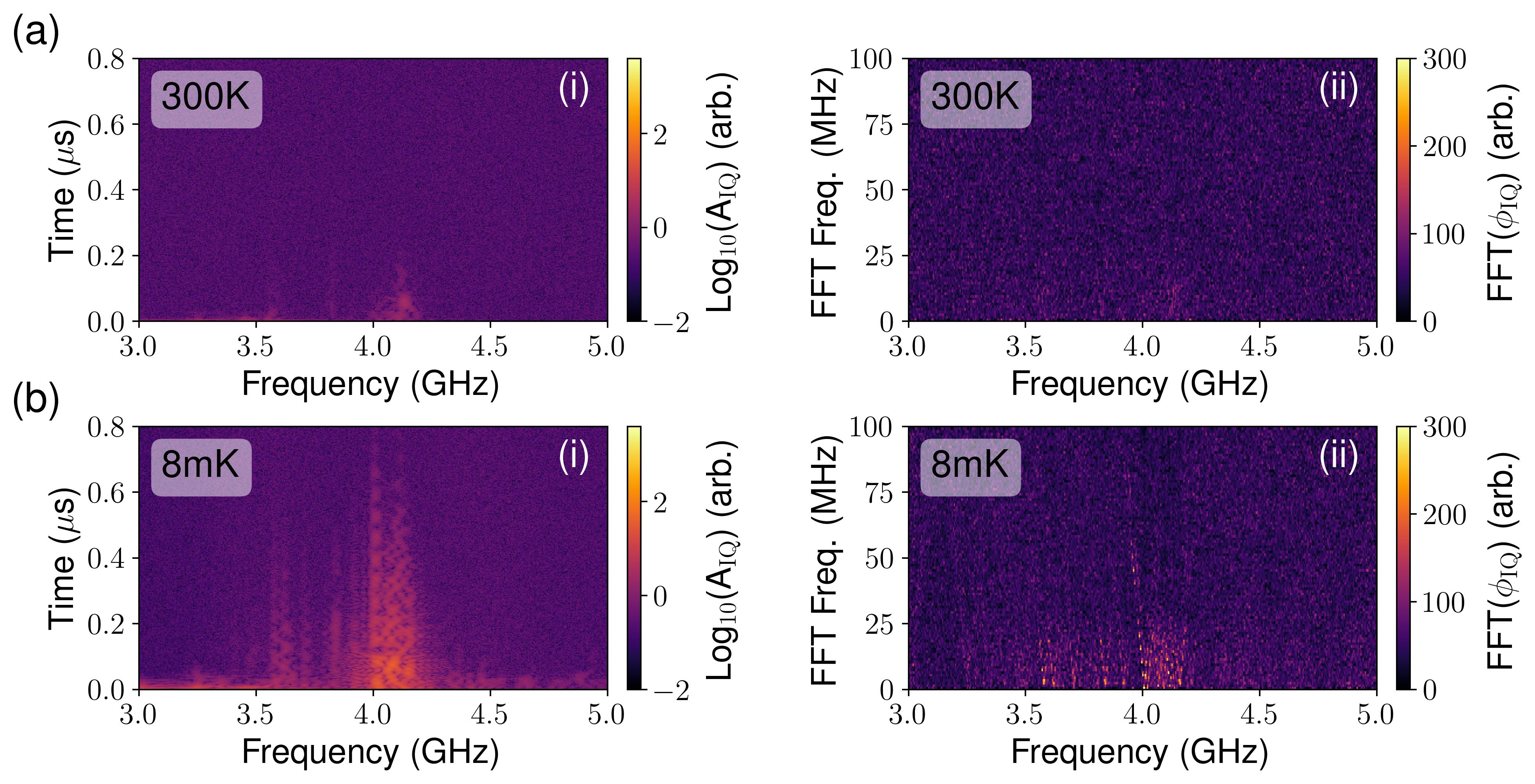}
\caption{BCTDS responses measured with the waveguide enclosed in the dilution refrigerator at room temperature (a) and at 8~mK (b). The room-temperature response is short-lived and spectrally sparse, whereas the cryogenic response is denser and more pronounced.
}
\label{fig:room_cryo_compare}
\end{figure*}

In the photoresist measurement shown below, in Appendix~\ref{app:mode_leakage}, Fig.~\ref{fig:room_cryo_compare}(b), we calibrated drive and readout powers, which allows us to estimate the transient power emitted from the waveguide. After accounting for the estimated attenuation and gain along the readout line, the initial ringdown power at a representative frequency of $4~\mathrm{GHz}$ is approximately $-65~\mathrm{dBm}$ ($P_{\mathrm{wg}}\sim 3\times10^{-10}~\mathrm{W}$) at the waveguide output. This corresponds to a photon flux
\[
\Phi_\gamma \sim \frac{P_{\mathrm{wg}}}{hf}
\sim
\frac{3\times10^{-10}~\mathrm{W}}
{(6.626\times10^{-34}~\mathrm{J\,s})(4\times10^9~\mathrm{s^{-1}})}
\sim 10^{14}~\mathrm{s^{-1}}.
\]
Since the transient response decays over a characteristic timescale of order $\sim100~\mathrm{ns}$, this flux corresponds to roughly
\[
N_\gamma \sim \Phi_\gamma \tau \sim 10^7
\]
traveling photons emitted over the ringdown timescale. This estimate is intended only as an order-of-magnitude photon-number scale for the emitted transient field, not as a stored photon occupation in a resonant mode. We also note that this estimate is not necessarily inconsistent with the relatively small number of resolved V-shaped resonances. These features may represent collective TLS defect sub-ensembles rather than individual defects, while many additional TLS defects contribute to an unresolved background. Thus, the estimated $\sim 10^7$ emitted photons may reflect the cumulative response of a much larger microscopic TLS defect population.

\section{Mode Leakage and Cavity Resonance}
\label{app:mode_leakage}

The WR-229 waveguide is designed to mainly support the propagating TE$_{10}$ mode, which has a strong oscillating electric field near the cross section at the center of the waveguide. The sample is mounted in this region, allowing the microwave drive to couple efficiently to the sample. In the ideal closed-waveguide geometry, the simulated transmission is relatively flat and shows minimal resonant structure, indicating that the field remains confined to the waveguide and does not form significant standing waves or cavity-like modes (Fig.~\ref{fig:mode_leakage}(ai) and (bi)).

In the experimental geometry, a thin rectangular slot is made to allow the sample to be inserted. This slot creates an opening in the waveguide wall, marked by the red box in Fig.~\ref{fig:mode_leakage}(aii), and provides a possible leakage path for microwave fields. HFSS simulations show that this opening introduces resonant dips in $S_{21}$, as shown in Fig.~\ref{fig:mode_leakage}(bii). Near these resonances, part of the electromagnetic field can leak out of the waveguide and reflect within the surrounding metallic enclosure. In our setup, this enclosure is a brass can at the mixing chamber stage.

We illustrate this leakage mechanism using a simplified qualitative simulation in which the waveguide is placed inside a conductive cylindrical shell. The simulated electric field at 4.5 GHz is shown in Fig.~\ref{fig:mode_leakage}(c). For the closed waveguide, the field remains confined inside the waveguide and behaves as a well-defined propagating TE$_{10}$ mode (Fig.~\ref{fig:mode_leakage}(ci)). When the sample slot is included, the field can leak through the opening and reflect between the waveguide and the surrounding conductive enclosure (Fig.~\ref{fig:mode_leakage}(cii)).

\begin{figure*}[htpb!]
\centering
\includegraphics[width=\textwidth,keepaspectratio]{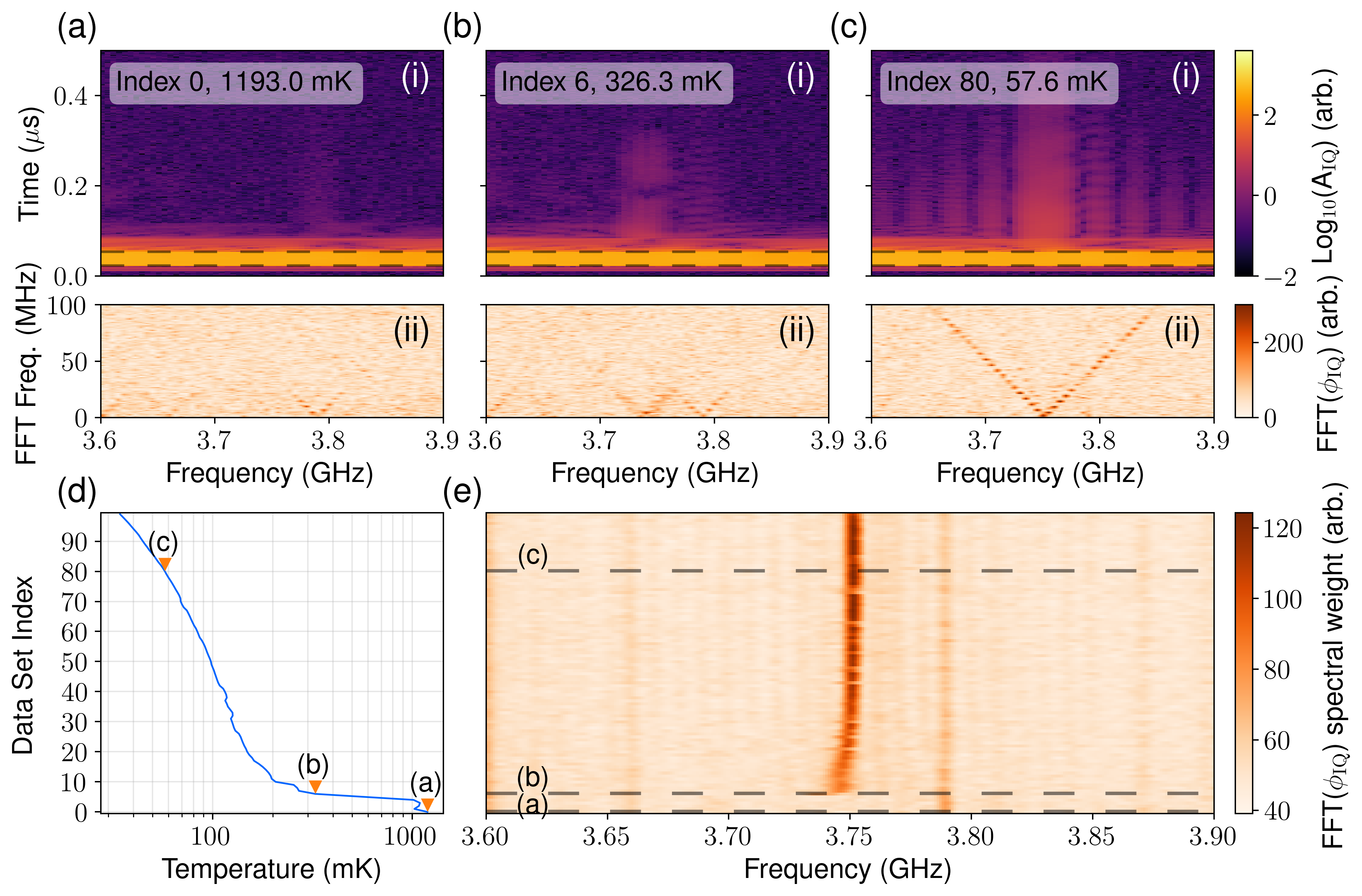}
\caption{Temperature dependence of BCTDS spectral features. We measure a single sapphire sample coated with Shipley 1813 photoresist and record back-to-back sweeps, indexed from 0 to 99, while cooling the dilution fridge from 1.2~K toward base temperature. (a)-(c), Representative transient spectra (i) and the FFT of phase signal (ii) at three representative instances: 1.2 K (index 0), 326 mK (index 6), and 57.6 mK (index 80), where we report the MXC temperature averaged across each sweep. (d) MXC temperature during the cooldown process. Slices (a-c) are marked with orange triangles. The actual sample temperature might be higher than the MXC sensor reading due to thermal conduction lag between the base plate and the insulating sapphire samples. (e) Compiled color map of the phase-V spectral weight across all sweeps (0–99) as a function of frequency, using a unit-slope V-feature fit method. We mark the representative slices (a-c) with dashed white lines. 
}
\label{fig:temperature_shift}
\end{figure*}

This simplified model is not intended to reproduce the exact mode structure of the experimental setup. In an actual experiment, the sample rack, SMA cables, neighboring devices, and deviations or finer details in the MXC-can geometry will all modify the frequencies and strengths of these cavity-like modes. The simulation instead provides a qualitative picture of how the sample slot can couple the waveguide mode to fields outside the waveguide. Consistent with this picture, small ringdown artifacts are observed even at room temperature, as shown in Fig.~\ref{fig:room_cryo_compare}(a). These features can persist as the temperature reduces, overlapping or mediating the response at base temperature. Importantly, we observe much denser and prolonged ring downs at base temperature (Fig.~\ref{fig:room_cryo_compare}(b)) that exhibit temperature and thermal cycle dependence, consistent with TLS defect behavior, as discussed in section~\ref{app:cooldown_compare} and Appendix~\ref{app:temp_shift}.

\section{Temperature Dependent Spectral Movement}
\label{app:temp_shift}

We observe a temperature-dependent shift in the transient response during a cooldown, as the dilution refrigerator cools from 1.2 K toward base temperature. We monitor the response over the 3.6--3.9 GHz frequency window and observe the sharp emergence of a resonant feature near 300 mK, corresponding to monitor index 6. We show representative data at 1.2 K, 326 mK, and 57 mK (monitor indices 0, 6, and 80, respectively) in Fig.~\ref{fig:temperature_shift}(a--c). Each panel shows the transient response (i) and the corresponding phase FFT (ii). To track the frequency of this feature, we estimate the V-shape spectral weight across the phase FFT spectrum. For each candidate center frequency, we average the FFT amplitude within slope-one diagonal regions corresponding to the expected V-shaped arms. This produces a frequency-dependent V-feature weight, whose peak identifies the location of the dominant response. We plot the extracted spectral weight for different monitor indices in Fig.~\ref{fig:temperature_shift}(e), with the corresponding MXC temperature reading shown in Fig.~\ref{fig:temperature_shift}(d). The feature not only emerges sharply near 300 mK, but also shifts in frequency as the system cools further. Below approximately 100 mK, it gradually converges to a stable frequency. This behavior is consistent with a TLS defect-related response. TLS defects with transition energies comparable to the thermal energy scale can change their occupation and interact with nearby thermally active defects during cooldown. These local configurational changes can modify the effective TLS defect environment and shift the observed resonance frequency. Once the temperature is sufficiently low, the thermally active degrees of freedom freeze out, and the resonance stabilizes.

\begin{figure}[htpb!]
\begin{centering}
\includegraphics[width=0.48\textwidth]{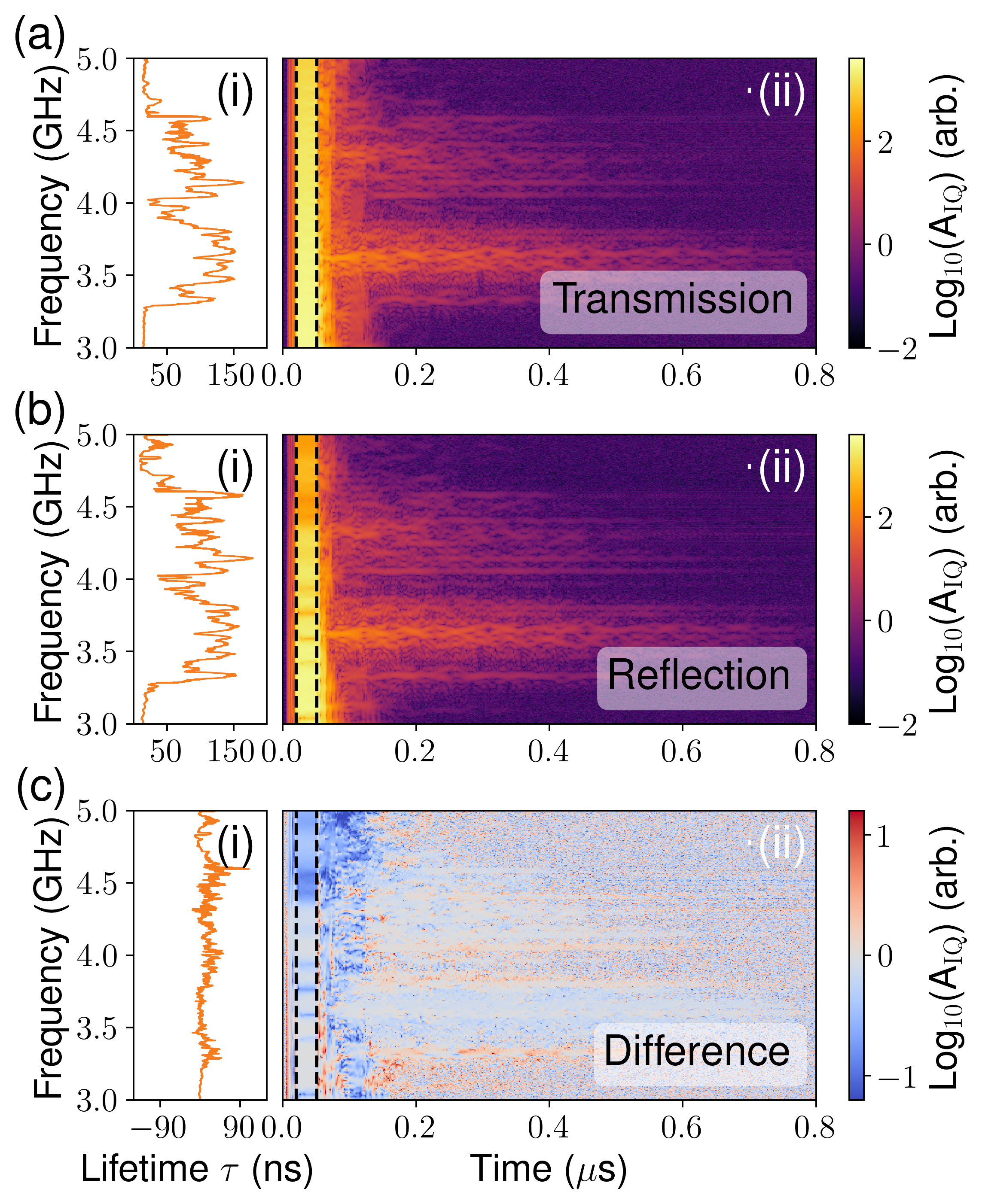}
\end{centering}
\caption{Dielectric response of Shipley 1813 photoresist on sapphire measured in transmission and reflection. We used the same samples as in the thermal-cycling experiment in Fig.~\ref{fig:cryo_compare}, but we add a cryogenic circulator (QCY-G0400801) to the waveguide input to measure both the transmitted (a) and reflected (b) signals. We also include a difference plot (c), computed as the difference between the $\log_{10}$ transient-response magnitudes, which shows a similarity in response from the transient region. In all panels, a short vertical white line marks the pulse bandwidth $\Delta f$, as in Fig.~\ref{fig:sample_compare}(d).}
\label{fig:transmission_reflection}
\end{figure}
%%%%%%%%%%%%%%%%%%%%%%%%%%%%%%%%%%%%%%%%%%%%%%%%%%%%%%%%%%%%%%%%%%

\section{Dielectric Response in Transmission and Reflection}
\label{app:T_R}
For ideal, lossless systems, transmission and reflection contain equivalent information as incident power must be conserved. In our experiment, where dissipation is present, reflection provides complementary information about dielectric loss during the pulsing region. We make measurements on the same samples as in Fig.~\ref{fig:cryo_compare} but place a cryogenic circulator at the waveguide input to measure both the transmission (Fig.~\ref{fig:transmission_reflection}(a)) and reflection (Fig.~\ref{fig:transmission_reflection}(b)) signals. In the transient region, emission from the dielectric samples is detected almost equally in the transmission and reflection channels, as the samples are positioned equidistant from the two antenna pins, and the readout circuits are identical. As a result, the collected signal shows minimal difference, as shown in Fig.~\ref{fig:transmission_reflection}(c). A short vertical white line is added to indicate pulse bandwidth $\Delta f$.

%%%%%%%%%%%%%%%%%%%%%%%%%%%%%%%%%%%%%%%%%%%%%%%%%%%%%%%%%%%%%%%%%%
\begin{figure}[htpb]
\includegraphics[width=0.48\textwidth]
{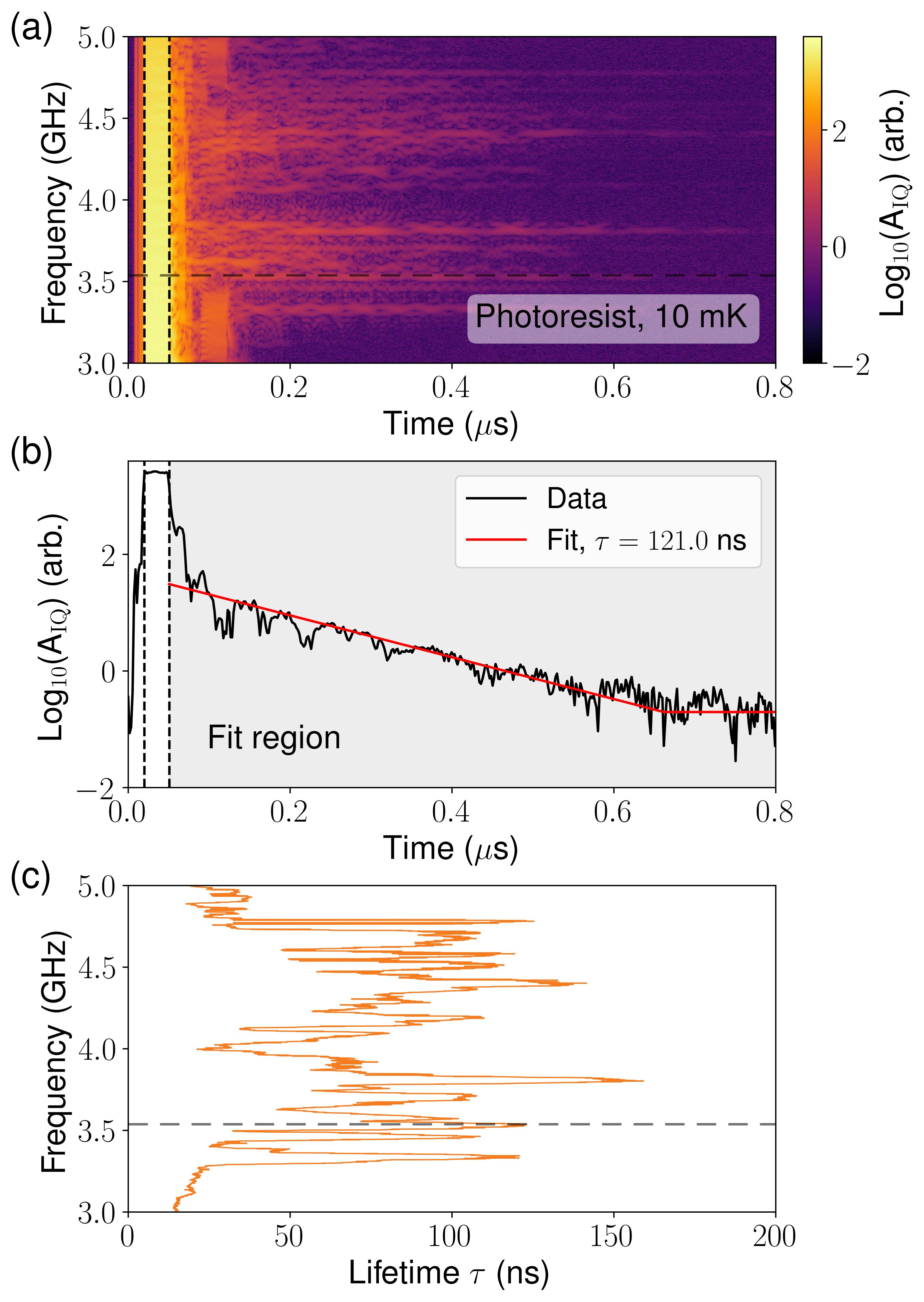}
\caption{Exponential decay analysis of the transient ringdown using a continuous two-piece linear fit in logarithmic amplitude.
(a) BCTDS spectroscopy result for a Shipley 1813 photoresist sample, shown as $\log_{10}(A_{\mathrm{IQ}})$ as a function of drive frequency and time.
(b) Representative trace at $3.54~\mathrm{GHz}$, marked by the black dashed line in (a). The black curve shows the full logarithmic-amplitude trace, including the drive pulse and subsequent transient response. The post-pulse transient region, shaded in gray, is fit in red using a continuous two-piece model consisting of an initial linear decay and a flat noise floor. The slope of the linear decay region is used to extract the ringdown lifetime $\tau$.
(c) Extracted ringdown lifetime $\tau$ as a function of drive frequency, obtained by applying the same fitting procedure across the full spectrum in (a).}
\label{fig:exp_fit}
\end{figure}
%%%%%%%%%%%%%%%%%%%%%%%%%%%%%%%%%%%%%%%%%%%%%%%%%%%%%%%%%%%%%%%%%%

\section{Additional Experiment and Analysis Details}
\subsection{Ringdown Exponential Fit}
\label{app:exp_fit}

The BCTDS ringdown exhibits beating, collapse, and revival features that reflect the collective dynamics of the driven TLS defect ensemble. Despite this nonmonotonic structure, the overall decay envelope follows an approximately exponential trend. To demonstrate the extraction of this decay rate, we fit the ringdown at each drive frequency for the Shipley sample response shown in Fig.~\ref{fig:exp_fit}(a). The extracted lifetimes are shown in Fig.~\ref{fig:exp_fit}(c). We demonstrate the exponential fitting process with a representative fit at 3.54~GHz, as shown in Fig.~\ref{fig:exp_fit}(b). For each trace, only the post-pulse transient is included in the fit. We model the decay envelope as
\begin{equation}
    A(t) = A_0 e^{-t/\tau},
\end{equation}
where $\tau$ is the characteristic decay time. In the analysis, we plot the logarithmic amplitude, $\log_{10}(A)$, for which an exponential decay becomes a linear trend,
\begin{equation}
    \log_{10} A(t) = \log_{10} A_0 - \frac{\log_{10}(e)}{\tau}t .
\end{equation}
At long times, the signal reaches a nearly constant noise floor. We therefore fit the logarithmic trace using a continuous two-piece linear model,
\begin{equation}
    y(t) =
    \begin{cases}
        kt+b, & t < t_c, \\
        kt_c+b, & t \geq t_c ,
    \end{cases}
\end{equation}
where $t_c$ marks the crossover to the noise floor. The fitted slope $k$ of the initial decay region is then converted to the exponential lifetime through
\begin{equation}
    \tau = -\frac{\log_{10}(e)}{k}.
\end{equation}
This procedure extracts the coarse decay envelope while reducing bias from the late-time noise floor. The remaining oscillatory structure around the fit reflects the underlying beating and revival dynamics of the TLS defect ensemble.

\subsection{HEMT Saturation}
\label{app:HEMT_sat}

\begin{figure}[hpbt!]
    \begin{centering}
    \includegraphics[width=0.48\textwidth]{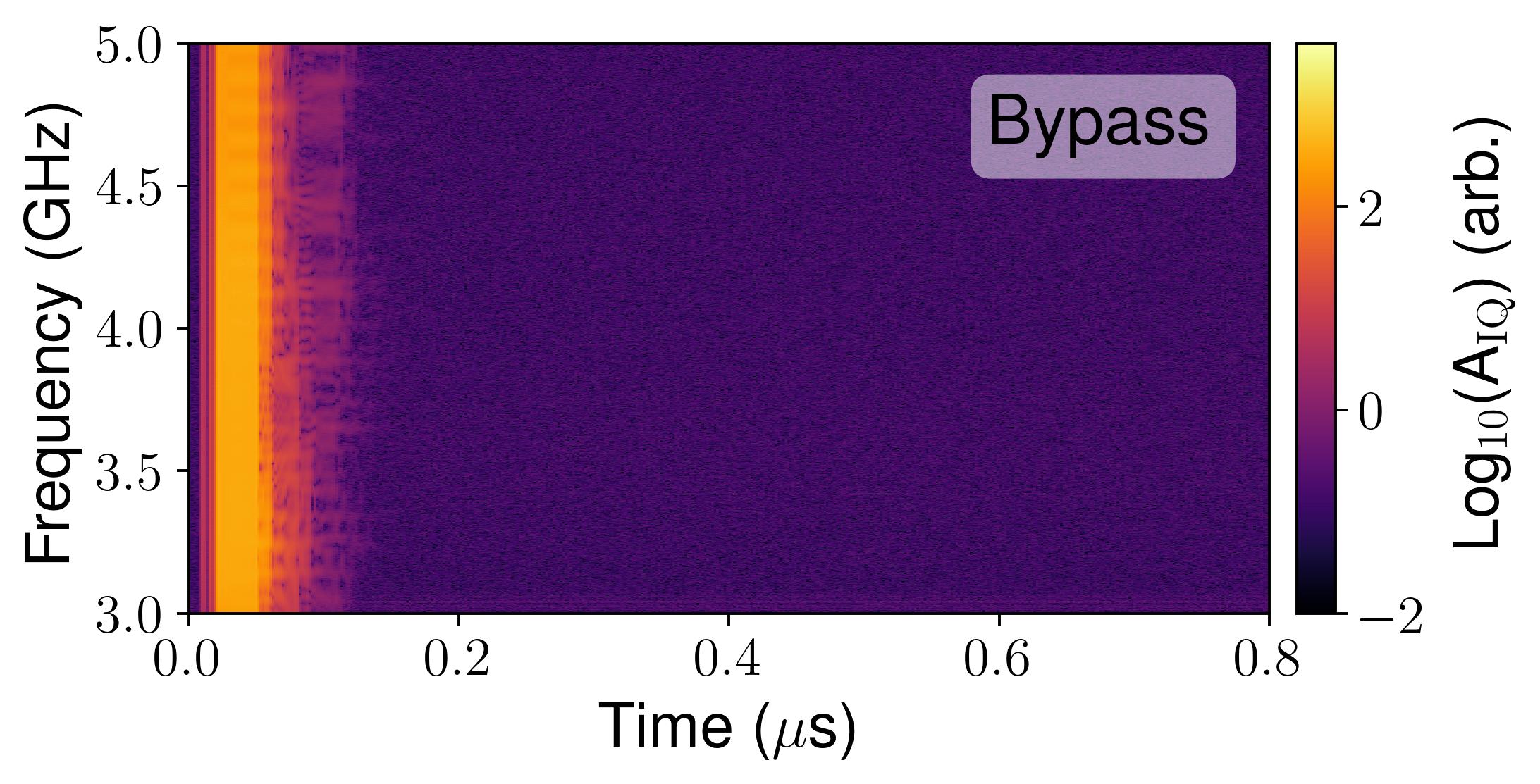}
    \end{centering}
    \caption{Bypass measurement at 8 mK using setup 2. The waveguide and sample are replaced with a direct SMA connection. The pulse briefly drives the HEMT (LNF-LNC4\_8C) amplifier into saturation, but the receiver chain quickly recovers and shows only a short-lived pulse artifact. This control measurement demonstrates that the long-lived ringdown observed in the sample measurements does not originate from HEMT saturation or nonlinear distortion in the receiver chain.}
    \label{fig:HEMT_bypass}
    \end{figure}

%%%%%%%%%%%%%%%%%%%%%%%%%%%%%%%%%%%%%%%%%%%%%%%%%%%%%%%%%%%%%%%%%%
\begin{figure}[htbp!]
\begin{centering}
\includegraphics[width=0.48\textwidth]{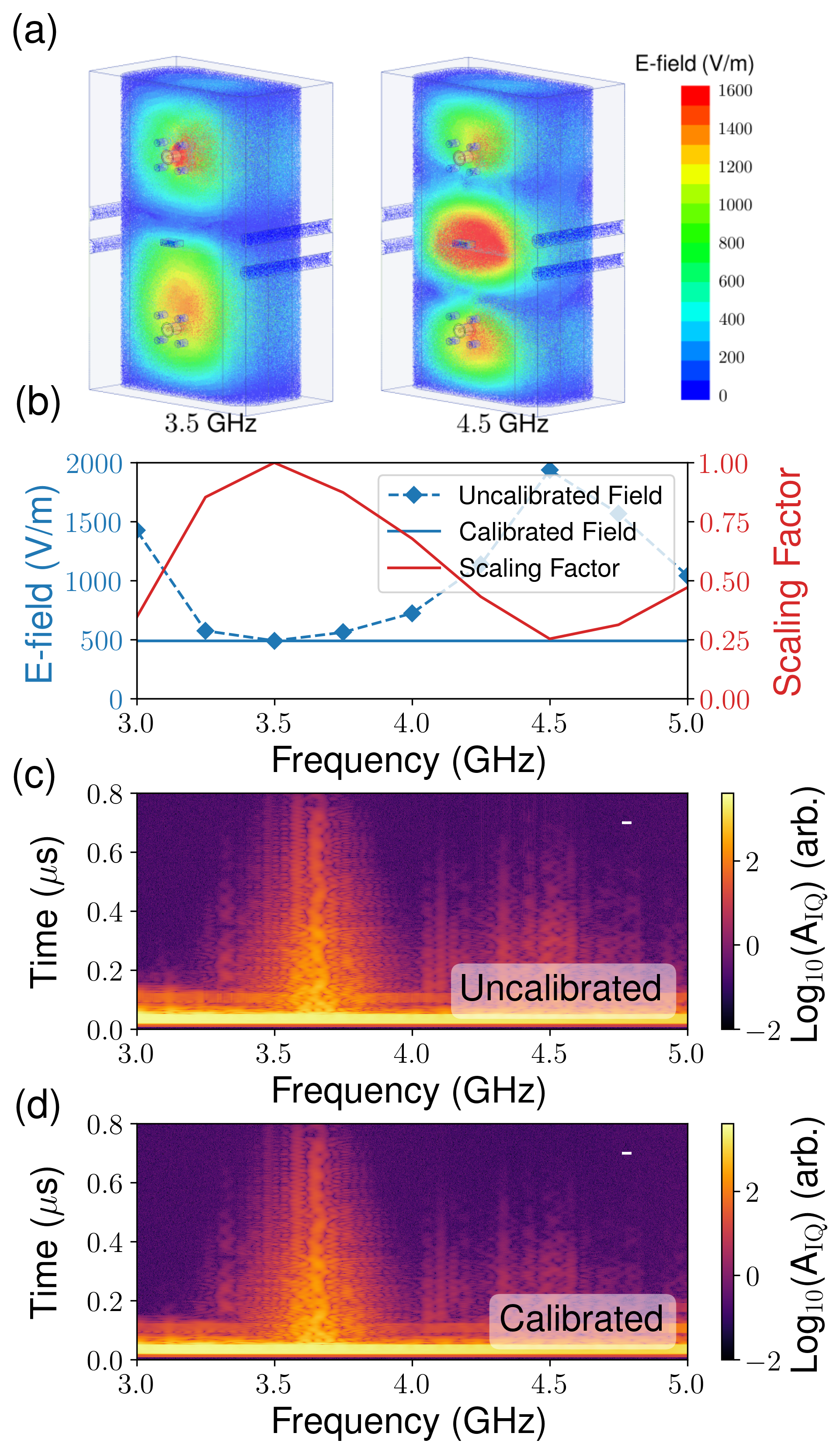}
\end{centering}
\caption{Electric field amplitude calibration at the sample plane. (a) HFSS simulation of the electric field distribution in the waveguide at 3 GHz and 5 GHz. The simulation applies an input power of 1 W at the SMA port to excite the dominant mode. The snapshot is taken for both frequencies when the oscillating electric field reaches its maximum amplitude at the sample plane. (b) Frequency-dependent electric field strength averaged over the sample plane and time (blue solid line). A scaling factor is applied to the pulse (red solid line) to achieve a uniform drive amplitude across the band, reducing the amplitude in regions with higher intrinsic field strength. The resulting average field profile is flattened (blue dashed line). (c)-(d) Comparison of transient spectra for a sapphire sample with AlO$_x$ deposit (same sample as Fig.~\ref{fig:sample_compare}(e) with a change in mounting configuration). We show the spectrum before (c) and after (d) field calibration. The calibrated spectrum shows reduced amplitude in previously over-driven regions and yields a more uniform interrogation of the ensembles of TLS defects across the full spectral range. A short horizontal white line is added to indicate pulse bandwidth $\Delta f$.}
\label{fig:e_field_calibration}
\end{figure}
%%%%%%%%%%%%%%%%%%%%%%%%%%%%%%%%%%%%%%%%%%%%%%%%%%%%%%%%%%%%%%%%%%

In this work, we intentionally apply a relatively strong drive to overcome the intrinsic localization of the TLS defects and induce a collective, synchronized response in the transient time window. As discussed in the wiring diagram Fig.~\ref{fig:fridge}, the estimated power delivered to the waveguide is approximately $-35~\mathrm{dBm}$. The HEMT amplifier, LNF-LNC4\_8C, has a specified input $P_{1\mathrm{dB}}$ of approximately $-55~\mathrm{dBm}$, so the HEMT is expected to briefly saturate during the drive pulse.

We therefore perform a bypass control measurement to test whether this saturation can generate the observed long-lived ringdown. In this measurement, the waveguide and sample are replaced by a direct SMA connection while keeping the rest of the receiver chain unchanged. The bypass measurement in Fig. \ref{fig:HEMT_bypass} shows only a short-lived pulse artifact immediately after excitation, with no long-lived ringdown. This result indicates that the long-lived transient response observed in the sample measurements originates from the sample response, rather than from HEMT saturation or nonlinear distortion in the receiver chain.

\subsection{Electric Field Amplitude Calibration}
\label{app:e_field_calib}

The waveguide design ensures that the transmitted signal is relatively constant over a bandwidth of approximately 3-6 GHz, shown in Fig.~\ref{fig:waveguide}(b). However, the samples under test are coupled in the middle of the waveguide, which can be frequency-dependent, as illustrated in Fig.~\ref{fig:e_field_calibration}(a). We leverage HFSS simulations to predict the average electric field amplitude at the sample plane over our frequency bandwidth , plotted in blue dots and lines in Fig.~\ref{fig:e_field_calibration}(b). To compensate for this variation, we apply a frequency-dependent scaling factor (red solid line). The impact of this calibration is evident when comparing the transient spectra of a sapphire sample with an AlO$_x$ deposit before (Fig.~\ref{fig:e_field_calibration}(c)) and after (Fig.~\ref{fig:e_field_calibration}(d)) applying the correction. Significantly, most of the key features in the data remain. Still, such calibrations can ensure uniform electric field densities as a function of frequency and may prove critical in future experiments to deduce properties of individual ensembles of TLS defects.

\section{Dielectric response of an ensemble of TLS defects}
\label{app:tls_response}

At low temperatures, the anomalous response of disordered insulating materials is commonly associated with ensembles of TLS defects with broadly distributed energy splittings and dipole couplings~\cite{Zeller1971}. In our experiment, a pulsed microwave field drives this ensemble directly inside a broadband superconducting waveguide, and the emitted post-pulse transient is measured by homodyne detection. This motivates a model that starts from the standard tunneling description, includes coupling to the applied electric field, and treats the subsequent transient dynamics within an open-system framework.

We begin with a single TLS defect in the position basis $\{|L\rangle,|R\rangle\}$, where the two localized configurations are separated by an asymmetry energy $\varepsilon_j$ and coupled by a tunneling matrix element $\Delta_j$. The corresponding Hamiltonian is
\begin{equation}
\hat H_j^{(p)}
=
\frac{1}{2}
\begin{pmatrix}
\varepsilon_j & \Delta_j\\
\Delta_j & -\varepsilon_j
\end{pmatrix}
=
\frac{\varepsilon_j}{2}\hat\sigma_z^{(p,j)}
+
\frac{\Delta_j}{2}\hat\sigma_x^{(p,j)}.
\label{eq:tls_position_basis_appendix}
\end{equation}
Diagonalization defines the energy splitting
\begin{equation}
E_j = \sqrt{\varepsilon_j^2+\Delta_j^2} = \hbar \omega_j,
\label{eq:tls_energy_splitting_appendix}
\end{equation}
and the mixing angle
\begin{equation}
\tan\theta_j = \frac{\Delta_j}{\varepsilon_j}.
\label{eq:tls_mixing_angle_appendix}
\end{equation}
In the energy basis, the static Hamiltonian becomes
\begin{equation}
\hat H_{0,j} = \frac{\hbar\omega_j}{2}\hat\sigma_z^{(j)}.
\label{eq:tls_energy_basis_appendix}
\end{equation}

In the position basis, an applied electric field modulates the relative energy of the two wells through the TLS dipole moment. Physically, this corresponds to periodically tilting the double-well potential and thereby driving transitions between the localized configurations. After transforming to the energy basis, the same coupling acquires both longitudinal and transverse components. The transverse component is proportional to $\hat\sigma_x^{(j)}$ and drives transitions between the energy eigenstates, while the longitudinal component modulates the level splitting.

For an ensemble of $N$ TLS defects, the collective polarization operator along the field direction can be written as
\begin{equation}
\hat P_x
=
\sum_{j=1}^{N}
p_j
\left(
\cos\theta_j\,\hat\sigma_z^{(j)}
+
\sin\theta_j\,\hat\sigma_x^{(j)}
\right),
\label{eq:collective_polarization_appendix}
\end{equation}
where $p_j$ is the dipole moment of the $j$th TLS defect projected along the field polarization axis. The driven Hamiltonian is then
\begin{equation}
\hat H(t)=\sum_{j=1}^{N}\hat H_{0,j}+\hat H_{\mathrm{drive}}(t),
\label{eq:driven_tls_hamiltonian_appendix}
\end{equation}
with
\begin{equation}
\hat H_{\mathrm{drive}}(t)=-\hat P_x E_x(t),
\label{eq:general_drive_hamiltonian_appendix}
\end{equation}
where $E_x(t)$ is the applied electric field. This expression provides the minimal driven extension of the standard tunneling model relevant for BCTDS.

In the weak-drive limit, the dielectric response can be expressed in terms of the dynamical susceptibility. The induced polarization is formally related to the applied field through
\begin{equation}
\langle \hat P_x(t)\rangle
=
\int_{-\infty}^{\infty}dt'\,\chi_{xx}(t-t')E_x(t'),
\label{eq:kubo_response_appendix}
\end{equation}
with retarded susceptibility
\begin{equation}
\chi_{xx}(t)
=
i\theta(t)\langle[\hat P_x(t),\hat P_x(0)]\rangle_0.
\label{eq:kubo_susceptibility_appendix}
\end{equation}
In frequency space,
\begin{equation}
\langle \hat P_x(\omega)\rangle=\chi_{xx}(\omega)E_x(\omega),
\qquad
\chi_{xx}(\omega)=\int_{-\infty}^{\infty}dt\,e^{i\omega t}\chi_{xx}(t),
\label{eq:susceptibility_frequency_appendix}
\end{equation}
and the dissipative part of the response is characterized by
\begin{equation}
\chi_{xx}''(\omega)\equiv \mathrm{Im}\,\chi_{xx}(\omega).
\label{eq:chi_imag_appendix}
\end{equation}
In our experiments, however, the drive can be strong and the response is not necessarily restricted to the linear regime. We therefore use the susceptibility formalism only as a qualitative framework for describing the dielectric response. The numerical treatment developed below instead follows the full time-dependent dynamics of the driven ensemble. Our main observable that we collect experimentally is the transient radiation emitted after the pulse, which directly reflects the time-dependent collective dipole response.

Within input-output theory, the detected homodyne signal is proportional to the radiated field and therefore to the collective polarization in the rotating frame,
\begin{equation}
E_{\mathrm{out}}(t)\propto \langle \hat P_x(t)\rangle e^{i\omega_d t}.
\label{eq:output_field_appendix}
\end{equation}
Experimentally, the in-phase and quadrature components, $I(t)$ and $Q(t)$, are recorded and combined into the complex output field. In the effective spin language used below, we consider the radiated field to be related to the collective coherence $\langle \sigma^+(t) \rangle \equiv \langle \sum_{j=1}^{N}\hat{\sigma}_+^{(j)} \rangle$, while the emitted intensity is related to $\langle \sigma^+\sigma^-(t) \rangle \equiv \langle (\sum_{j=1}^{N}\hat{\sigma}_+^{(j)})(\sum_{j=1}^{N}\hat{\sigma}_-^{(j)}) \rangle$. Accordingly, the measured amplitude and phase of the homodyne signal can be interpreted in terms of the transient collective dipole coherence prepared during the pulse and measured during the post-pulse ring-down.

\section{Driven Open-System Model and Collective Radiative Decay}
\label{app:lindblad_model}

To describe the transient dynamics following pulsed excitation, we extend the driven TLS defect framework introduced above to include effective interactions between defects and radiative decay into the broadband waveguide. In the energy basis, the interacting Hamiltonian is written as
\begin{equation}
\hat H(t)
=
\sum_{j=1}^{N}\frac{\hbar\omega_j}{2}\hat\sigma_z^{(j)}
+
\hat H_{\mathrm{drive}}(t)
+
\sum_{i<j}J_{ij}\hat\sigma_x^{(i)}\hat\sigma_x^{(j)},
\label{eq:interacting_tls_hamiltonian_appendix}
\end{equation}
where $\omega_j$ is the bare frequency of the $j$th TLS defect and $J_{ij}$ denotes the effective coupling between TLS defects. The interaction term is written in the same basis used to describe the driven dipole response and provides a minimal model for coherent coupling within the ensemble.

For the square cosine pulse used in the simulations, we retain the transverse component of the dipole coupling and write
\begin{equation}
\hat H_{\mathrm{drive}}(t)
=
\sum_{j=1}^{N} \Omega(t)\cos(\omega_d t)\,\hat\sigma_x^{(j)},
\label{eq:drive_term_square_pulse_appendix}
\end{equation}
with envelope
\begin{equation}
\Omega(t)=
\begin{cases}
\Omega, & 0\le t\le \tau_{\mathrm{p}},\\
0, & \text{otherwise}.
\end{cases}
\label{eq:pulse_envelope_appendix}
\end{equation}
Here $\tau_{\mathrm{p}}$ denotes the pulse duration. This form keeps the part of the drive that directly induces transitions between the energy eigenstates and captures the driven dynamics most relevant for the post-pulse response considered here.

The open-system dynamics is described by the Lindblad master equation
\begin{equation}
\frac{d\rho}{dt}
=
-i[\hat H(t),\rho]
+
\Gamma
\left(
2\hat S_-\rho \hat S_+
-
\hat S_+\hat S_-\rho
-
\rho \hat S_+\hat S_-
\right),
\label{eq:lindblad_collective_appendix}
\end{equation}
where
\begin{equation}
\hat S_{\pm}=\sum_{j=1}^{N}\hat\sigma_{\pm}^{(j)},
\label{eq:collective_ladder_appendix}
\end{equation}
and $\Gamma$ is the collective radiative decay rate. The collective jump operator $\hat S_-$ is motivated by the broadband waveguide, which couples many TLS defects to the same radiative environment and provides a channel for collective emission.

The observables used throughout the numerical analysis are
\begin{equation}
\langle \sigma^+\sigma^- \rangle
=
\langle \hat S_+\hat S_- \rangle,
\label{eq:collective_population_appendix}
\end{equation}
and
\begin{equation}
\phi \approx \arg\!\left(\langle \sigma^+ \rangle\right)
=
\arg\!\left(\langle \hat S_+ \rangle\right).
\label{eq:collective_phase_appendix}
\end{equation}
These quantities provide an interpretative framework for relating the simulated TLS defect dynamics to the homodyne observables measured experimentally. The output field is proportional to the collective dipole radiation of the driven ensemble, so its amplitude is related to the emitted transient intensity, while its phase tracks the collective coherence of the radiated field. In this sense, $\langle \sigma^+\sigma^- \rangle$ serves as a proxy for the transient emitted power, and $\phi$ captures the phase evolution of the collective emission.

\begin{figure}[htpb!]
\centering
\includegraphics[width=0.5\textwidth]{figures/floquet_quasienergies_large_fonts.png}
\caption{
Floquet interpretation of the post-pulse transient response in a driven two-spin system. Numerical results are obtained from Eq.~\ref{eq:lindblad_collective_appendix} for a square cosine pulse with dipole-dipole coupling $J/2\pi = 20~\mathrm{MHz}$, pulse duration $\tau_{\mathrm{p}}=100~\mathrm{ns}$, collective decay rate $\Gamma=1.0~\mathrm{MHz}$, and bare transition frequencies $\omega_{1}/2\pi=3.75~\mathrm{GHz}$ and $\omega_{2}/2\pi=3.82~\mathrm{GHz}$, indicated by vertical dashed lines. Left and right columns correspond to $\Omega/2\pi = 0.01~\mathrm{GHz}$ and $0.1~\mathrm{GHz}$, respectively. (a) Integrated post-pulse tail area versus drive frequency. (b) Collective population $\langle \sigma^+\sigma^- \rangle$ as a function of time and drive frequency; the horizontal dashed line marks the end of the pulse. (c) Magnitude of the Fourier spectrum $|\mathrm{FFT}(\phi)|$ of the transient phase response, showing V-shaped branches centered near the bare transition frequencies of the individual TLS defects. (d) Lowest Floquet quasi-energies computed during the driving interval. More pronounced post-pulse transients occur at drive frequencies where the Floquet spectrum develops small quasienergy differences.}
\label{fig:floquet_quasienergies}
\end{figure}

\section{Floquet Picture Generated During the Pulse}
\label{app:floquet_formalism}

The driven-dissipative model introduced above describes the transient response after the pulse, but during the pulse itself the Hamiltonian is approximately periodic at the drive frequency. This makes Floquet theory a useful interpretive framework for understanding how the drive prepares the state that later evolves during the post-pulse ring-down. In particular, it provides a qualitative connection between the measured transient response and the quasienergy structure generated during the finite driving interval.

For a time-periodic Hamiltonian,
\begin{equation}
\hat H(t+T)=\hat H(t),
\qquad
T=\frac{2\pi}{\omega_d},
\label{eq:periodic_hamiltonian_appendix}
\end{equation}
the evolution over one period is described by the one-period propagator
\begin{equation}
\hat U(T)=\mathcal{T}\exp\!\left[-i\int_0^T \hat H(t)\,dt\right],
\label{eq:one_period_propagator_appendix}
\end{equation}
where $\mathcal{T}$ denotes time ordering. The eigenvalues of $\hat U(T)$ can be written as
\begin{equation}
\hat U(T)|u_\alpha(0)\rangle=e^{-iQ_\alpha T}|u_\alpha(0)\rangle,
\label{eq:floquet_eigenvalue_appendix}
\end{equation}
which defines the Floquet quasienergies $Q_\alpha$. Equivalently, solutions of the Schr\"odinger equation take the Floquet form
\begin{equation}
|\psi_\alpha(t) \rangle = e^{-iQ_\alpha t}|u_\alpha(t)\rangle,
\label{eq:floquet_state_appendix}
\end{equation}
where $|u_\alpha(t)\rangle$ is periodic with period $T$. Expanding the periodic part in harmonics,
\begin{equation}
|u_\alpha(t)\rangle = \sum_{m=-\infty}^{\infty}e^{-im\omega_d t}|u_\alpha^{(m)}\rangle,
\label{eq:floquet_harmonics_appendix}
\end{equation}
leads to the standard Fourier-space Floquet eigenvalue problem
\begin{equation}
\sum_m \mathcal H_{nm}|u_\alpha^{(m)}\rangle
=
Q_\alpha |u_\alpha^{(n)}\rangle,
\label{eq:sambe_eigenproblem_appendix}
\end{equation}
with matrix elements
\begin{equation}
\mathcal H_{nm}=H_{n-m}+m\omega_d\delta_{nm},
\qquad
H_k=\frac{1}{T}\int_0^T dt\,e^{ik\omega_d t}\hat H(t).
\label{eq:floquet_matrix_appendix}
\end{equation}

The central point is that the drive generates dressed Floquet states during the pulse, and the post-pulse dynamics can inherit quasienergy differences between the populated Floquet components. When two quasienergies $Q_\alpha$ and $Q_\beta$ become nearly degenerate, the difference $|Q_\alpha-Q_\beta|$ is small, producing slow phase evolution after the drive is turned off. If these Floquet states are appreciably populated during the pulse and overlap with the measured collective observables, this slow phase evolution can produce enhanced low-frequency spectral weight and more prominent ring-downs. Quasienergy near-degeneracy alone is therefore not sufficient, namely the states must also be populated by the drive and contribute to $\langle \sigma^+\sigma^- \rangle$ or $\langle \sigma^+\rangle$.

\begin{figure*}[htpb!]
\centering
\includegraphics[width=\textwidth,keepaspectratio]{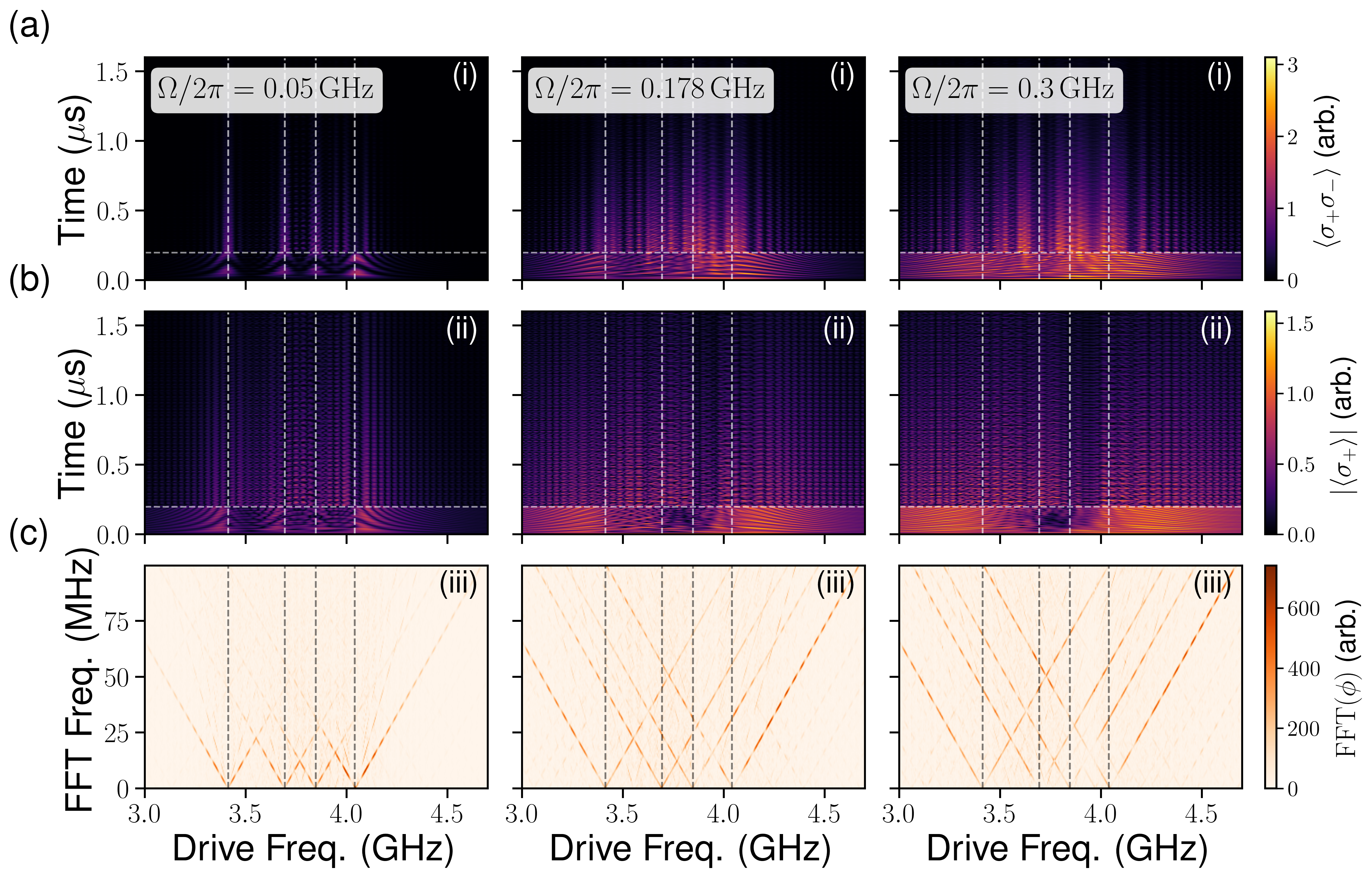}
\caption{Drive-amplitude dependence of the transient response in an interacting four-spin system. Numerical results are obtained from Eq.~\ref{eq:lindblad_collective_appendix} for a single disorder realization of an all-to-all coupled network with bare transition frequencies drawn from $\omega_j/2\pi \in[3.0,5.0]~\mathrm{GHz}$, corresponding to $\omega_1/2\pi = 3.693~\mathrm{GHz}$, $\omega_2/2\pi = 3.412~\mathrm{GHz}$, $\omega_3/2\pi = 3.846~\mathrm{GHz}$, and $\omega_4/2\pi = 4.039~\mathrm{GHz}$, constant coupling rates $J_{ij}/2\pi = 25~\mathrm{MHz}$, collective decay rate $\Gamma=2.0~\mathrm{MHz}$, and pulse duration $\tau_{\mathrm{p}}=200~\mathrm{ns}$. Vertical dashed lines mark the bare transition frequencies. Columns correspond to $\Omega/2\pi=0.05~\mathrm{GHz}$, $0.178~\mathrm{GHz}$, and $0.3~\mathrm{GHz}$. (a) Collective population $\langle \sigma^+\sigma^- \rangle$ as a function of time and drive frequency. (b) Magnitude of the collective coherence $|\langle \sigma^+ \rangle|$ under the same conditions. (c) Magnitude of the Fourier spectrum $|\mathrm{FFT}(\phi)|$ of the transient phase response. Increasing the drive amplitude broadens the frequency range over which the ensemble responds and strengthens the multi-branch V-shaped spectral features.}
\label{fig:simulation_driving_amplitude}
\end{figure*}

This picture also connects naturally to Landau--Zener--St\"uckelberg physics. During the pulse, the dressed energy levels are repeatedly driven through avoided crossings. Each passage can partially transfer amplitude between dressed branches, and the phase accumulated between passages,
\begin{equation}
\Phi_{\mathrm{St}} \sim \int Q_\alpha(t)\,dt,
\label{eq:stuckelberg_phase_appendix}
\end{equation}
controls the interference between the resulting amplitudes. We use this Floquet--Landau--Zener--St\"uckelberg picture as a qualitative framework for understanding how the finite-duration pulse prepares the post-pulse transient.

This interpretation is illustrated in Fig.~\ref{fig:floquet_quasienergies} for a driven two-spin system. Figures~\ref{fig:floquet_quasienergies}(a-i) and \ref{fig:floquet_quasienergies}(a-ii) show the integrated post-pulse tail area for weak and stronger driving. At weak drive, the response is localized near the bare transition frequencies, marked by the dashed vertical lines. At stronger drive, the tail area extends over a wider range of drive frequencies and develops additional structure around the resonances. This is consistent with stronger drive-induced dressing, where off-resonant frequencies contribute more visibly to the post-pulse response.

The corresponding time-domain maps in Fig.~\ref{fig:floquet_quasienergies}(b) show the post-pulse evolution of $\langle \sigma^+\sigma^- \rangle$. For a weak drive, the response is weak and concentrated near resonance. For a stronger drive, the response persists over a broader range of drive frequencies and shows clear oscillations after the pulse is turned off. In this model, these oscillations arise from coherent evolution between states populated during the driven interval.

Figure~\ref{fig:floquet_quasienergies}(c) shows the Fourier spectrum of the phase, $|\mathrm{FFT}(\phi)|$. The stronger-drive case displays V-shaped branches centered near the bare transition frequencies, consistent with the phase-FFT operating principle described in Sec.~\ref{sec:op_principle}: after the pulse, the phase evolves at rates set by detunings between the drive and the relevant TLS defect frequencies. Figure~\ref{fig:floquet_quasienergies}(d) shows the Floquet quasienergies during the pulse. The strongest post-pulse response occurs near regions where the quasienergy curves approach or hybridize, indicating that the finite drive can prepare coherent superpositions of dressed states. These simulations provide an interpretative framework for the experiment: the transient ring-down reflects coherences prepared during the pulse, while the phase-FFT structure tracks the corresponding detuning-dependent phase evolution of the collective emission.

\begin{figure*}[htpb!]
\centering
\includegraphics[width=\textwidth,keepaspectratio]{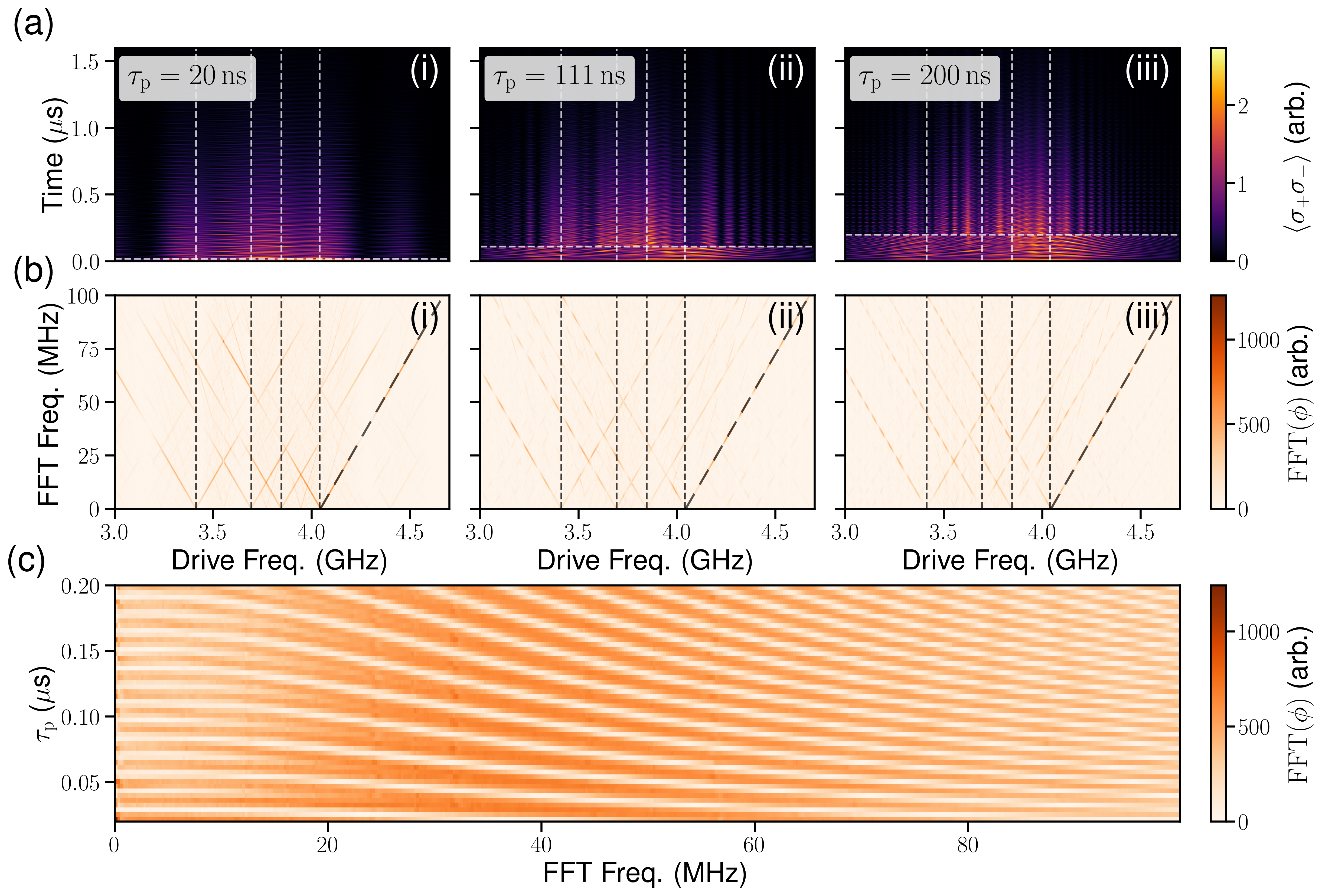}
\caption{
Pulse-duration dependence of transient spectral sharpening in an interacting four-spin system. Numerical results are obtained from Eq.\,\ref{eq:lindblad_collective_appendix} for a single disorder realization with the same bare transition frequencies as in Fig.~\ref{fig:simulation_driving_amplitude}, constant coupling rate $J_{ij}/2\pi = 25~\mathrm{MHz}$, collective radiative decay rate $\Gamma=2.0~\mathrm{MHz}$, and drive amplitude $\Omega/2\pi=0.2~\mathrm{GHz}$. Vertical dashed lines mark the bare transition frequencies. Columns (i)--(iii) correspond to $\tau_{\mathrm{p}}=20~\mathrm{ns}$, $111~\mathrm{ns}$, and $200~\mathrm{ns}$. (a) Collective population $\langle \sigma^+\sigma^- \rangle$ versus time and drive frequency; the horizontal dashed line marks the end of the pulse. (b) Magnitude of the Fourier spectrum $|\mathrm{FFT}(\phi)|$ of the transient phase response. Longer pulses sharpen the V-shaped branches without shifting their vertices. The dashed guide indicates the spectral arm used for the analysis. (c) Fourier amplitude evaluated along the selected branch as a function of pulse duration and Fourier frequency. The oscillatory fringe pattern reflects phase accumulation during the finite pulse and interference between transient components in the post-pulse response.}
\label{fig:simulation_pulse_width_ringdown_fft}
\end{figure*}

\section{Numerical Demonstration of the Drive-Amplitude Dependence}
\label{app:amplitude_dependence}

We next examine how the transient response changes as the drive amplitude is increased in an interacting four-spin realization. This provides a numerical illustration of the Floquet picture introduced above. At weak drive, the response is concentrated near isolated resonances. As the drive amplitude increases and becomes comparable to the typical detuning between nearby defects, more dressed transitions participate in the driven response, leading to broader spectral features and stronger post-pulse dynamics.

These trends are shown in Fig.~\ref{fig:simulation_driving_amplitude}. Figure~\ref{fig:simulation_driving_amplitude}(a) presents the collective population $\langle \sigma^+\sigma^- \rangle$ for three increasing drive amplitudes. At low amplitude, the response is confined to narrow frequency regions near the bare transition frequencies of the individual TLS defects. As the amplitude increases, the response broadens in drive frequency and the post-pulse oscillations become more visible. The same trend appears in Fig.~\ref{fig:simulation_driving_amplitude}(b), where the collective coherence $|\langle \sigma^+ \rangle|$ develops stronger beating at larger drive amplitudes. The broader and overlapping spectral features indicate that off-resonant TLS defect components contribute more strongly to the driven response.

The corresponding phase spectra in Fig.~\ref{fig:simulation_driving_amplitude}(c) show multi-branch V-shaped features that become stronger and more extended with increasing drive amplitude. Within the Floquet interpretation, these branches arise from detuning-dependent phase evolution associated with dressed-state components populated during the pulse. Stronger driving increases Floquet mixing and allows more quasienergy differences to contribute to the measured phase response.

The spectra also develop interference structure along and between the V-shaped branches. This structure reflects the coherent superposition of transient components prepared during the pulse. In the complementary Landau--Zener--St\"uckelberg picture, the modulations are associated with phase accumulation between repeated passages through nearby avoided crossings. After the pulse, this accumulated phase appears as modulation in the Fourier-domain response.

Taken together, Fig.~\ref{fig:simulation_driving_amplitude} shows that increasing the drive amplitude broadens the participating frequency range, strengthens the post-pulse transient, and enhances interference features in the phase-FFT response. These trends are consistent with stronger drive-induced dressing and spectral mixing in the interacting TLS defect model.

\begin{figure*}[htpb!]
\centering
\includegraphics[width=\textwidth,keepaspectratio]{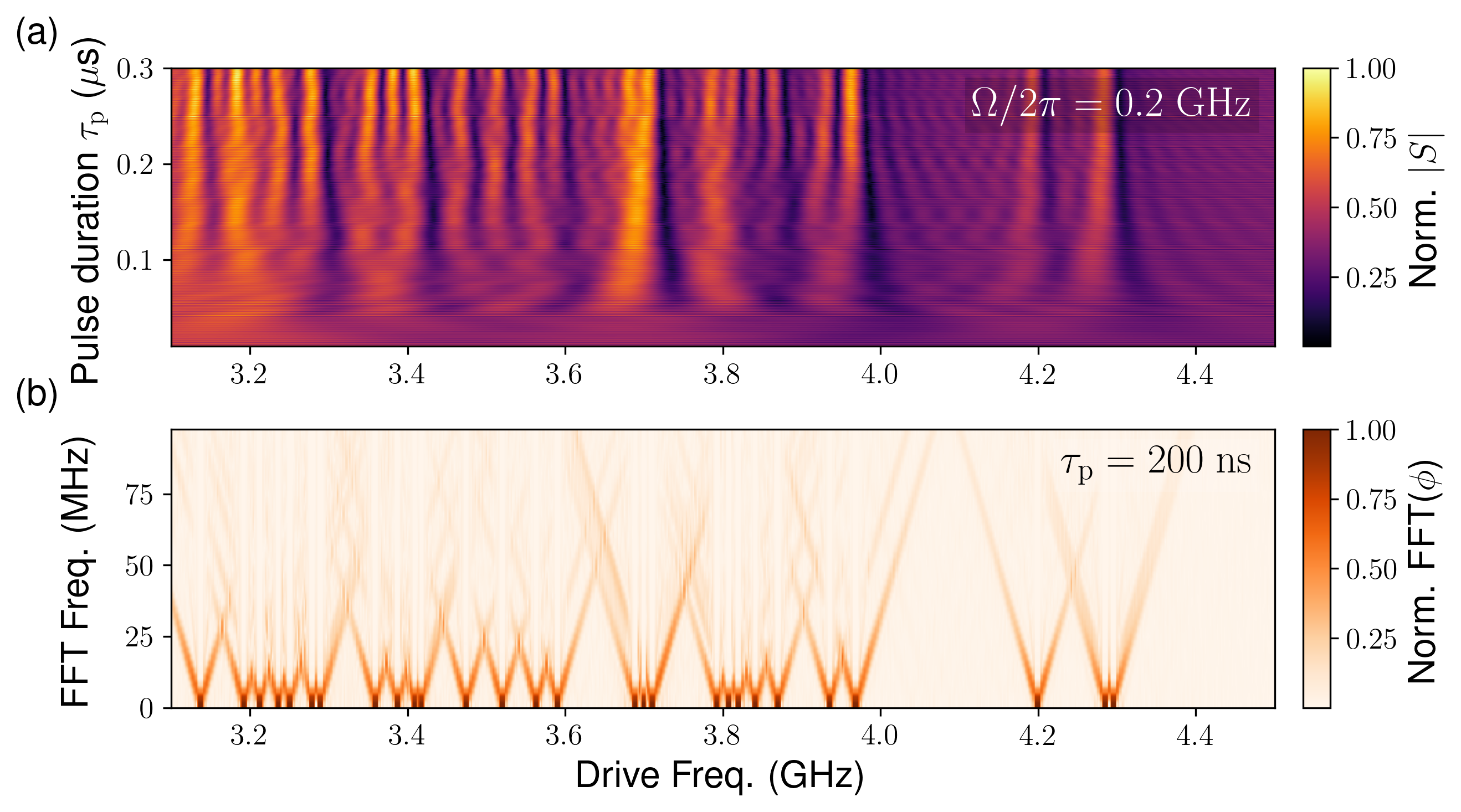}
\caption{
Weak-drive analytical limit for a non-interacting disordered TLS defect ensemble. Results are shown for a single realization of $N=30$ TLS defects with bare frequencies $\omega_j/2\pi\in[3.1,4.5]~\mathrm{GHz}$ and drive amplitude $\Omega/2\pi=0.2~\mathrm{GHz}$. (a) Magnitude of the collective coherence $|S|$ from Eq.\,\ref{eq:ensemble_time_integrated_coherence_appendix} as a function of drive frequency and pulse duration. Longer pulses narrow the response and sharpen the interference pattern. (b) Magnitude of the Fourier spectrum $|\mathrm{FFT}(\phi)|$ of the phase response for a fixed pulse duration $\tau_{\mathrm{p}}=200~\mathrm{ns}$. The resulting branches are centered at the bare TLS defect frequencies, showing that the basic V-shaped spectral features already emerge in the weak-drive, non-interacting limit.}
\label{fig:analytics_multispin}
\end{figure*}

\section{Pulse-Duration Dependence: Spectral Sharpening and Interference Fringes}
\label{app:pulse_dependence}

The pulse duration, $\tau_{\mathrm{p}}$, controls the spectral selectivity of the excitation and the phase accumulated during the drive. Short pulses have broad spectral bandwidth and can excite several nearby transitions, producing diffuse Fourier features. Longer pulses are more spectrally selective and allow phase to accumulate over a longer interval, which sharpens the branches observed in the post-pulse phase response.

This behavior is shown in Fig.~\ref{fig:simulation_pulse_width_ringdown_fft} for the same interacting four-spin model used above. The time-domain maps of $\langle \sigma^+\sigma^- \rangle$ in Fig.~\ref{fig:simulation_pulse_width_ringdown_fft}(a) show that increasing $\tau_{\mathrm{p}}$ produces more clearly resolved post-pulse oscillations. The corresponding Fourier spectra in Fig.~\ref{fig:simulation_pulse_width_ringdown_fft}(b) show that the V-shaped branches narrow systematically as $\tau_{\mathrm{p}}$ is increased. The branch vertices remain fixed near the underlying transition frequencies, while the pulse duration mainly controls the spectral resolution and contrast of the branches.

To examine the interference structure, we track selected spectral arms indicated by the black dashed lines in Fig.~\ref{fig:simulation_pulse_width_ringdown_fft}(b). We then evaluate the Fourier amplitude along these trajectories as a function of $\tau_{\mathrm{p}}$. The resulting maps in Fig.~\ref{fig:simulation_pulse_width_ringdown_fft}(c) show oscillatory fringes. These fringes are consistent with coherent phase accumulation during the pulse, namely along a given spectral branch, the relevant detuning changes with drive frequency, and the accumulated phase changes with $\tau_{\mathrm{p}}$. In the Landau--Zener--St\"uckelberg picture, the same modulation can be viewed as interference between amplitudes generated by repeated passages through dressed-state avoided crossings.

Within this framework, Fig.~\ref{fig:simulation_pulse_width_ringdown_fft}(c) shows that the spectral branches are sensitive to the time spent under the drive. This is consistent with the Floquet interpretation developed above, where the finite pulse populates dressed states and their quasienergy differences determine the structure of the subsequent transient response.

\section{Analytical Description in the Weak-Drive, Non-Interacting Limit}
\label{app:analytics}

To isolate the basic origin of the arms of the V-spectral features and their sharpening with pulse duration, we consider a weak-drive, non-interacting limit. This analytical treatment is not intended to reproduce the full interacting transient response. Instead, it provides a minimal framework for understanding how finite-time coherent driving generates the phase structure that appears in the post-pulse spectrum. 

For a single TLS defect with bare frequency $\omega_j$, we consider
\begin{equation}
\hat H_j(t)
=
\frac{\hbar\omega_j}{2}\hat\sigma_z
+
\Omega \cos(\omega_d t)\hat\sigma_x,
\label{eq:single_tls_driven_hamiltonian_appendix}
\end{equation}
with $\Omega \ll \omega_j$ and $\omega_d \sim \omega_j$.
In the \(|g\rangle, |e\rangle\) basis, the state is
\[
|\psi(t)\rangle = c_g(t)\,|g\rangle + c_e(t)\,|e\rangle,
\]
where \(c_g(t)\) and \(c_e(t)\) are the ground- and excited-state amplitudes. The coupled time-dependent Schr\"odinger equations for \(c_e^{(j)}(t)\) and \(c_g^{(j)}(t)\) can be reduced to a single exact second-order differential equation for \(c_e^{(j)}\):
\begin{equation}
\ddot c_e^{(j)}
+
\left[
\frac{\omega_j^2}{4}
+
\frac{\Omega^2}{2}
+
\frac{\Omega^2}{2}\cos(2\omega_d t)
+
\frac{\omega_j\Omega}{2}\cos(\omega_d t)
\right]
c_e^{(j)}
=0.
\end{equation}
This form makes the periodically driven structure explicit and leads to a generalized Mathieu, or Hill-type, equation \cite{Zuo1994,Sarkar2021}.

Introducing the dimensionless variable \(\theta=\omega_d t\), the equation becomes
\begin{equation}
\frac{d^2 c_e^{(j)}}{d\theta^2}
+
\big[
A_j
+
B_j\cos\theta
+
C_j\cos 2\theta
\big]
c_e^{(j)}
=0,
\end{equation}
with
\[
A_j =
\frac{\omega_j^2/4+\Omega^2/2}{\omega_d^2},
\qquad
B_j =
\frac{\omega_j\Omega}{2\omega_d^2},
\qquad
C_j =
\frac{\Omega^2}{2\omega_d^2}.
\]
In the weak-drive regime, \(\Omega\ll\omega_j\), the \(C_j\) term can be neglected to leading order. Since the coefficients are periodic, Floquet theory allows the solution to be written as
\begin{equation}
c_e^{(j)}(\theta)
=
e^{-i\epsilon_j\theta}
\sum_{n=-\infty}^{\infty}
u_n^{(j)} e^{-in\theta},
\end{equation}
where \(u_n^{(j)}\) are Fourier coefficients and \(\epsilon_j\) is the Floquet quasienergy.
Substituting this expansion into the Hill equation gives the recurrence relation
\begin{align}
(\epsilon_j+n)^2 u_n^{(j)}
&=
A_j u_n^{(j)}
+
\frac{B_j}{2}
\left(
u_{n-1}^{(j)}+u_{n+1}^{(j)}
\right)
\nonumber\\
&\quad+
\frac{C_j}{2}
\left(
u_{n-2}^{(j)}+u_{n+2}^{(j)}
\right).
\end{align}
In the weak-drive limit, keeping only the nearest-neighbor couplings (\(C_j\simeq0\)) reduces the recurrence relation to
\begin{equation}
(\omega_j-n\omega_d)u_n^{(j)}
=
\frac{\Omega}{2}
\left(
u_{n-1}^{(j)}+u_{n+1}^{(j)}
\right),
\end{equation}
whose standard solution is given by the Bessel-function expansion
\begin{equation}
u_n^{(j)}
=
J_n\!\left(\frac{\Omega}{\omega_d}\right).
\end{equation}

In the weak to intermediate driving limit, the excited-state amplitude can therefore be expanded in Floquet harmonics as
\begin{equation}
c_e^{(j)}(t)
\approx
\sum_{n=-\infty}^{\infty}
J_n\!\left(\frac{\Omega}{\omega_d}\right)
e^{-i(\omega_j-n\omega_d)t}.
\label{eq:single_tls_floquet_harmonics_appendix}
\end{equation}
Each harmonic evolves with detuning \(\omega_j-n\omega_d\). Integrating this coherence over a pulse of duration $\tau_{\mathrm{p}}$ gives
\begin{equation}
S_j(\tau_{\mathrm{p}})
\approx
\sum_{n=-\infty}^{\infty}
J_n\!\left(\frac{\Omega}{\omega_d}\right)
\frac{1-e^{-i(\omega_j-n\omega_d)\tau_{\mathrm{p}}}}
{\omega_j-n\omega_d}.
\label{eq:single_tls_time_integrated_coherence_appendix}
\end{equation}
For a non-interacting TLS defect ensemble, the total response is
\begin{equation}
S(\tau_{\mathrm{p}})
\approx
\sum_{j=1}^{N_{\mathrm{TLS}}}
\sum_{n=-\infty}^{\infty}
J_n\!\left(\frac{\Omega}{\omega_d}\right)
\frac{1-e^{-i(\omega_j-n\omega_d)\tau_{\mathrm{p}}}}
{\omega_j-n\omega_d}.
\label{eq:ensemble_time_integrated_coherence_appendix}
\end{equation}

Equation~\ref{eq:ensemble_time_integrated_coherence_appendix} captures two key effects. The denominator selects drive frequencies near the bare TLS defect frequencies and their Floquet harmonics, while the finite-time factor controls both spectral sharpening and phase accumulation during the pulse. In practice, we use $|S(\tau_{\mathrm{p}})|$ to construct the pulse-duration maps and define
\begin{equation}
\phi(\omega_d)=\arg\!\left[S(\tau_{\mathrm{p}}^\ast,\omega_d)\right]
\label{eq:phase_from_analytic_response_appendix}
\end{equation}
for a fixed pulse duration $\tau_{\mathrm{p}}^\ast$, from which we compute $|\mathrm{FFT}(\phi)|$.

The results are summarized in Fig.~\ref{fig:analytics_multispin}. Figure~\ref{fig:analytics_multispin}(a) shows $|S|$ for a non-interacting disordered ensemble as a function of drive frequency and pulse duration. As in the numerical simulations above, the response sharpens systematically as $\tau_{\mathrm{p}}$ increases. Figure~\ref{fig:analytics_multispin}(b) shows the corresponding Fourier amplitude $|\mathrm{FFT}(\phi)|$ for a fixed long pulse. The same V-shaped branches observed in the Lindblad simulations already appear in this weak-drive limit. This shows that the basic V-shaped features and pulse-duration spectral selectivity can arise from finite-time coherent driving of a disordered non-interacting ensemble. Interactions and collective decay are therefore not required to generate the basic branch structure, although they can modify the contrast, linewidths, and interference patterns of the full transient response.

\begin{figure}[h]
\centering
\includegraphics[width=0.5\textwidth]{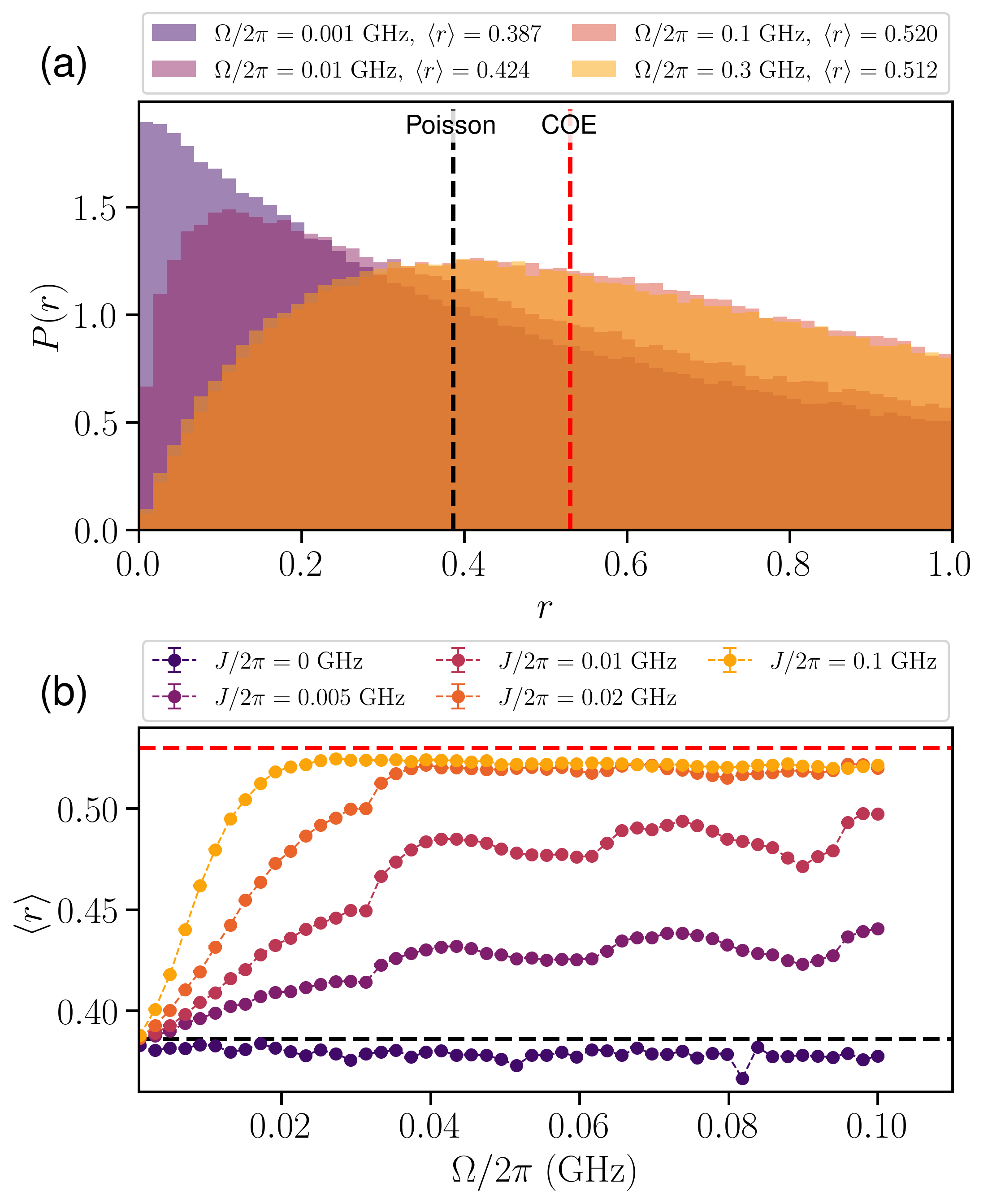}
\caption{
Floquet level statistics for the driven interacting spin model. The adjacent-gap ratio $r$ is computed from the ordered quasienergy spectrum of the Floquet propagator and averaged over disorder realizations. The dashed black and red reference lines mark the Poisson and circular orthogonal ensemble (COE) benchmarks, respectively. (a) Distribution $P(r)$ for selected drive amplitudes, with the corresponding mean values $\langle r\rangle$ indicated in the legend. Increasing the drive amplitude shifts the distribution from Poisson-like toward COE-like behavior. (b) Disorder-averaged mean gap ratio $\langle r\rangle$ as a function of drive amplitude for several interaction strengths $J/2\pi$. In the non-interacting limit, the Floquet spectrum remains close to Poisson statistics, whereas finite interactions induce clear level repulsion and drive the system toward the COE regime.}
\label{fig:level_statistics}
\end{figure}

\section{Perspective on the Role of Interactions and Drive-Induced Delocalization}
\label{app:level_statistics_discussion}

The transient simulations discussed above describe the post-pulse dynamics of a driven disordered TLS defect ensemble, but they do not directly address how interactions reorganize the underlying Floquet spectrum. To address this question, we analyze quasienergy level statistics following Ref.~\cite{Ponte2015}, which distinguishes weakly hybridized Floquet spectra from strongly mixed driven regimes.

Following the stroboscopic construction of Ref.~\cite{Ponte2015}, we consider a piecewise-driven disordered interacting spin model for which the one-period propagator can be written as a product of simple unitaries. In our implementation, the Hamiltonian is
\begin{equation}
H_1= J\sum_{i<j}\sigma_i^z\sigma_j^z+\sum_i h_i\sigma_i^z,
\qquad
H_2=H_1+\Omega\sum_i \sigma_i^x,
\label{eq:fc_ising_hamiltonians_appendix}
\end{equation}
with $h_i$ drawn from a uniform disorder distribution. The corresponding Floquet operator is
\begin{equation}
U_F=e^{-iH_2T/2}e^{-iH_1T/2},
\label{eq:stroboscopic_floquet_operator_appendix}
\end{equation}
which can be diagonalized directly for many disorder realizations and parameter values. From the ordered quasienergies $\epsilon_n$ of $U_F$, we compute the adjacent-gap ratio
\begin{equation}
r_n=\frac{\min(\delta_n,\delta_{n+1})}{\max(\delta_n,\delta_{n+1})},
\qquad
\delta_n=\epsilon_{n+1}-\epsilon_n,
\label{eq:gap_ratio_appendix}
\end{equation}
whose disorder-averaged value distinguishes different spectral regimes. For uncorrelated quasienergies, one expects $\langle r\rangle \approx 0.386$ (Poisson), while a delocalized time-reversal-symmetric Floquet spectrum approaches $\langle r\rangle \approx 0.53$ (COE) \cite{Ponte2015}.

The results are shown in Fig.~\ref{fig:level_statistics}. In Fig.~\ref{fig:level_statistics}(a), the distribution $P(r)$ shifts from Poisson-like to COE-like as the drive amplitude is increased. In Fig.~\ref{fig:level_statistics}(b), the disorder-averaged mean $\langle r\rangle$ remains near the Poisson value in the non-interacting limit, while finite interactions drive it towards the COE benchmark. Within this model, increasing drive and interactions therefore promotes stronger Floquet level repulsion and a more hybridized spectrum.

In the context of BCTDS, this calculation should be viewed as a perspective as a spectral diagnostic rather than a direct model of the measured transient response. Following Ref.~\cite{Ponte2015}, we use a simplified stroboscopic spin model, rather than the full driven-dissipative TLS defect Hamiltonian, as a computationally efficient proxy to assess how increasing drive and coupling strength promote Floquet spectral hybridization. Within this framework, the resulting level statistics are consistent with a localization-to-delocalization crossover that may also be relevant in the context of the driven TLS defect ensemble that we explore here.

\section{Localization-to-Delocalization Crossover and Connection to BCTDS}
\label{app:numerical_outlook}

The level-statistics analysis complements the driven-dissipative TLS defect simulations developed in the previous sections. The Lindblad calculations were introduced to describe the experimentally relevant BCTDS observables, namely the transient amplitude and phase response. By contrast, the Floquet analysis probes how disorder, drive, and interactions modify the spectral structure generated during the drive.

This perspective is again motivated by Ref.~\cite{Ponte2015}, where quasienergy statistics were used to characterize localization and delocalization in periodically driven disordered systems. Here, we adapt the same stroboscopic idea to a driven interacting spin model that is simple enough to survey over many disorder realizations and parameter values. For that reason, Fig.~\ref{fig:level_statistics} should be interpreted as evidence for a localization-to-delocalization \emph{crossover} in the model, rather than as a direct phase-diagram statement for the experimental TLS defect ensemble.

This distinction is important. In BCTDS, the experimentally accessible quantities are the transient homodyne amplitude and phase, which are described in our framework by $\langle \sigma^+\sigma^- \rangle$ and $\phi \approx \arg(\langle \sigma^+ \rangle)$. The level-statistics calculation does not replace that description. Instead, it provides a compact spectral measure showing that larger $J$ and $\Omega$ push the model toward a more strongly mixed Floquet regime. This is consistent with a picture in which more hybridized modes contribute to the structured post-pulse emission.

Taken together, the transient simulations and the Floquet level-statistics analysis provide a consistent numerical picture for BCTDS, namely, the pulsed drive generates a structured nonequilibrium Floquet response, and increasing interactions and drive strength enhance the degree of spectral hybridization and multi-photon processes that could contribute to the observed collective transient dynamics.

% \bibliography{apssamp}% Produces the bibliography via BibTeX.
% \bibliographystyle{apsrev4-2}
% \bibliography{references}


\end{document}